# Penalized MM Regression Estimation with $L_\gamma$ Penalty: A Robust Version of Bridge Regression

Olcay Arslan[1]

Department of Statistics, Ankara University, 06100 Tandogan, Ankara, Turkey

The bridge regression estimator generalizes both ridge regression and LASSO estimators. Since it minimizes the sum of squared residuals with a $L_\gamma$ penalty, this estimator is typically not robust against outliers in the data. There have been attempts to define robust versions of the bridge regression method, but while these proposed methods produce bridge regression estimators robust to outliers and heavy-tailed errors, they are not robust against leverage points. We propose a robust bridge regression estimation method combining MM and bridge regression estimation methods. The MM bridge regression estimator obtained from the proposed method is robust against outliers and leverage points. Furthermore, for appropriate choices of the penalty function, the proposed method is able to perform variable selection and parameter estimation simultaneously. Consistency, asymptotic normality, and sparsity of the MM bridge regression estimator are achieved. We propose an algorithm to compute the MM bridge regression estimate. A simulation study and a real data example are provided to demonstrate the performance of the MM bridge regression estimator for finite sample cases.

*Keywords*: Bridge regression, Lasso, MM estimator, Penalized regression, Ridge regression, Robust regression, Variable selection.

## 1   Introduction

Consider the linear regression model

$$y_i = \alpha + \mathbf{x}_i^T \beta + \varepsilon_i, \quad i = 1,2,3,\ldots,n, \qquad (1)$$

where $y_i \in R$ is the response variable; $\mathbf{x}_i = (x_{i1}, x_{i2}, \ldots, x_{ip})^T$ is the $p-$dimensional vector of the explanatory variables; $\beta = (\beta_1, \beta_2, \ldots, \beta_p)^T$ is the vector of regression parameters in $R^p$; and $\varepsilon_i$'s are the iid random errors with zero mean, $\sigma^2$ variance and the distribution $F$. Without loss of generality, we assume that $\alpha = 0$ and consider the model

$$y_i = \mathbf{x}_i^T \beta + \varepsilon_i, \quad i = 1,2,3,\ldots,n. \qquad (2)$$

The regression equation given in (2) can also be written in matrix notation as

---

[1]  E-mail: oarslan@ankara.edu.tr; Tel.: +90 312 212 67 20/1210; Fax:+90 312 2233202; Address: Department of Statistics, Faculty of Science, Ankara University, 06100 Tandogan, Ankara, Turkey

$$Y = X\beta + \varepsilon,$$

where $X_{n \times p}$ is the design matrix, $Y$ is the response vector, and $\varepsilon$ is the vector of $\varepsilon_i$. Throughout this study, $\beta_0 = (\beta_{01}, \beta_{02}, ..., \beta_{0p})^T$ denotes the true parameter vector and $\Omega \subset R^p$ will denote the parameter space. We assume that $\Omega$ is compact and $\beta_0$ is in the interior of $\Omega$.

One way of estimating the unknown parameter vector $\beta$ is to minimize the following penalized least squares (LS) objective function

$$Q_n(\beta) = \sum_{i=1}^{n}(y_i - \mathbf{x}_i^T \beta)^2 + \lambda_n \sum_{j=1}^{p} |\beta_j|^{\gamma}, \qquad (3)$$

where $\gamma > 0$, and $\lambda_n$ is the penalty regularization parameter. This method of penalized least squares, which was introduced by Frank and Friedman (1993), is called the bridge regression estimation method and the function

$$L_{\gamma}(\beta) = \sum_{j=1}^{p} |\beta_j|^{\gamma} \qquad (4)$$

is the bridge penalty function. The bridge regression is introduced as a generalization of the ridge regression ($\gamma = 2$). It also includes the LASSO regression method ($\gamma = 1$) introduced by Tibshirani (1996). Frank and Friedman (1993) explain the roles of the parameters $\lambda_n$ and $\gamma$ as follows (see page 124): *the parameter $\lambda_n$ regulates the strength of the penalty, and the parameter $\gamma$ controls the degree of preference for the true regression coefficients $\beta$ to align with the original variable directions in the predictor space. A value $\gamma = 2$ yields a rotationally invariant penalty function expressing no preference for any particular direction. The case $\gamma > 2$ results in a prior that supposes that the true coefficient vector is more likely to be aligned in directions oblique to the variable axes, whereas for $\gamma < 2$ it is more likely to be aligned with the variable axes.*

$L_{\gamma}(\beta)$ is a convex function of $\beta$ for $\gamma \geq 1$ and a non-convex function of $\beta$ for $0 < \gamma < 1$. Note that $\gamma = 0$ yields the entropy penalty function (Antoniandis and Fan, 2001). When $0 < \gamma \leq 1$, the bridge regression method attempts to shrink the small regression coefficients to exact zeros, and hence selects important variables. Therefore, the bridge regression method provides a way of combining parameter estimation and variable selection in a single minimization problem. Further, for the case $0 < \gamma < 1$, the resulting estimator will be nearly unbiased. That is, the estimator is unbiased for large values of the unknown parameter vector $\beta$ (Fan and Li, 2001). When $\gamma > 1$, the bridge regression method shrinks the regression coefficients, but does not provide variable selection. It was shown by Knight and Fu (2000) and Liu et al. (2007) that, for larger values of $\gamma$, the shrinkage increases with the magnitude of the regression parameters being estimated. The suggested value of $\gamma$ is between $0$ and $2$.

Several researchers have investigated the properties of the bridge regression estimator after its original definition by Frank and Friedman (1993), for example Fu (1998), Knight and Fu (2000), Fan and Li (2001), Liu et al. (2007), Huang et al. (2008), Li and Yu (2009), Armagan (2009), Caner

(2009) and Park and Yoon (2011). The asymptotic properties of the bridge regression estimator were explored by Knight and Fu (2000) and Huang et al. (2008). In these papers, consistency, asymptotic normality and sparsity of the bridge estimator were established under some appropriate conditions. It was shown that when $0 < \gamma < 1$, the bridge regression method correctly identifies zero coefficients and the estimators of the nonzero coefficients are asymptotically normal and have the oracle property. The case $\gamma = 1$ gives the LASSO estimators, and its asymptotic properties are well established. For the case $\gamma > 1$, the consistency and asymptotic normality of the bridge regression have also been studied.

In general, to obtain the bridge regression estimator that minimizes the objective function given in (3) requires handling a nonlinear optimization problem, which is not easy to solve. Fu (1998) proposed a general approach to solve the bridge regression minimization problem for the case $\gamma \geq 1$. For the case $\gamma > 0$, the local quadratic approximation (LQA) and the local linear approximation (LLA) algorithms are used to minimize the objective function given in (3) (Fan and Li, 2001; Hunter and Li, 2005, Liu et al., 2007; Zou and Li, 2008; Parker and Yoon, 2011). Note that the LQA and the LLA algorithms are examples of the majorization-minimization algorithm, which is an extension of the well-known EM algorithm (see Hunter and Li, 2005 and Zou and Li, 2008). This identification guarantees the convergence of the LQA and the LLA algorithms.

Since the bridge regression estimation method is based on the LS method, it is sensitive to outliers and/or heavy-tailed errors. Concerning the LASSO and the ridge regression estimators, which are the special cases of the bridge regression (BR) estimator, several approaches have been proposed to make them robust against heavy-tailed errors and/or outliers in the response, for example Owen (2007), Wang and Leng (2007), and Xu and Ying (2010). Maronna (2011), Arslan (2012), and Alfons et al. (2013) have considered the robust ridge regression, the weighted LAD-LASSO regression, and the sparse LTS regression, respectively, to make the ridge regression estimator and the LASSO estimator robust against the leverage points (outliers in the explanatory variables). However, there are few proposals to make the BR regression estimator robust for the case $\gamma > 0$. Li and Yu (2009) proposed robust and sparse BR based on a generalized Huber function to make the BR estimator robust against heavy-tailed errors or outliers in the response. Li et al. (2011) considered the non-concave penalized M estimation method in sparse and high-dimensional linear regression models to carry out parameter estimation and variable selection simultaneously.

All attempts to define a robust bridge regression estimator are based on the M estimation method. As a result, the resulting robust bridge estimators can only handle outliers in response, and cannot be robust against leverage points. The purpose of the present paper is to combine the high breakdown point regression estimation method and the BR estimation method to obtain a robust bridge regression estimator that is resistant to outliers in the response, leverage points, and/or heavy-tailed errors. To achieve this we combine the MM regression method, introduced by Yohai (1987), and the bridge regression method. The MM regression method produces estimators that are resistant to outliers in any direction and the bridge regression method will either do the variable selection for $\gamma \leq 1$ or shrink the estimators for $\gamma > 1$. The proposed method and the estimator obtained from this method will be called the MM bridge regression (MM-BR) estimation method and the MM-BR estimator, respectively.

In Section 2 we introduce the MM-BR estimator and study its asymptotic properties. In Section 3 we propose an algorithm based on the LQA to compute the MM-BR estimate. In Section 4 we provide a simulation study and a real data example to demonstrate the performance of the MM-BR estimation method in terms of model selection and parameter estimation. Finally, we discuss our outcomes and conclusions in Section 5. The proofs of the propositions stated in Section 2 are given in the Appendix.

## 2 Robust Bridge Regression Estimation

We first review the MM regression estimation method proposed by Yohai (1987), then define the robust bridge regression estimation method based on the MM regression method.

### 2.1 MM Regression Estimation

The MM regression estimation method yields an estimator that is resistant to outliers in the data (outliers in the explanatory and the response variables) and are efficient under normality. The MM method is based on two different $\rho$ functions, $\rho_0$ and $\rho_1$, to determine the breakdown point and the efficiency. These functions should be bounded and have the following properties (see Yohai 1987):

A1: (i) $\rho(0) = 0$, (ii) $\rho(-x) = \rho(x)$, (iii) $\rho(x)$ is continuous, (iv) $\sup \rho(x) = a < \infty$, (v) if $0 \leq x \leq y$ $\rho(x) \leq \rho(y)$, (vi) if $\rho(x) < a$ and $0 \leq x < y$, then $\rho(x) < \rho(y)$.

A2: Let $\rho_1(x) \leq \rho_0(x)$ and $\sup \rho_1(x) = \sup \rho_0(x) = a$.

The MM regression estimator can be obtained as follows. We start with an initial estimator, $\hat{\beta}_s$, for the parameter vector $\beta$. This initial estimator should be robust with high breakdown point, but does not necessarily need to be efficient. Using $\hat{\beta}_s$, we compute the initial residuals $r_i(\hat{\beta}_s) = y_i - \mathbf{x}_i^T \hat{\beta}_s$, for $i = 1,2,3,...,n$. Then, find a scale M estimator $\hat{\sigma}_s$ for the initial residuals $r_i(\hat{\beta}_s)$ by solving the following M estimating equation for the scale parameter

$$\frac{1}{n}\sum_{i=1}^{n}\rho_0\left(\frac{r_i(\hat{\beta}_s)}{\sigma}\right) = 0.5. \tag{5}$$

Finally, we find the absolute minimum of

$$l_n(\beta) = \sum_{i=1}^{n}\rho_1\left(\frac{r_i(\beta)}{\hat{\sigma}_s}\right) \tag{6}$$

to obtain the MM estimator, ($\hat{\beta}_{MM}$), for the parameter vector $\beta$, where $r_i(\beta) = y_i - \mathbf{x}_i^T \beta$. If $\rho_1$ has a derivative $\Psi_1(t) = \rho_1'(t)$, then the MM regression estimator $\hat{\beta}_{MM}$ of $\beta$ will be any solution of the estimating equation

$$\sum_{i=1}^{n}\Psi_1\left(\frac{r_i(\beta)}{\hat{\sigma}_s}\right)\mathbf{x}_i = \mathbf{0}, \tag{7}$$

which verifies

$$l_n(\hat{\beta}_{MM}) \leq l_n(\hat{\beta}_s). \tag{8}$$

The choice of the $\rho$ functions is also important to have efficiency and high breakdown point. Yohai (1987), Maronna et al. (2006), and Maronna (2011) use the bisquare $\rho$-function

$$\rho(r) = \min\{1, 1-(1-r^2)^3\}, r \in R, \tag{9}$$

with $\rho_0(r) = \rho(r/c_0)$ and $\rho_1(r) = \rho(r/c_1)$. To have $\rho_1 \leq \rho_0$ we must choose $c_0 \leq c_1$. In this paper we will use the same $\rho$ function.

See Yohai (1987), Maronna et al. (2006), and Maronna and Yohai (2010) for further details and properties of the MM regression method.

### 2.2 MM-BR Estimation

Consider the regression model given in (2). Let $\hat{\beta}_s$ be the initial robust estimator defined in Subsection 2.1 and let $\hat{\sigma}_s$ be the scale M estimator for the residuals $r_i(\hat{\beta}_s)$, $i = 1, 2, , n$. Then, the MM-BR estimator is defined by

$$\hat{\beta}_{MM-BR} = \arg\min_{\beta} L_n(\beta), \tag{10}$$

where

$$L_n(\beta) = \hat{\sigma}_s^2 \sum_{i=1}^{n} \rho_1\left(\frac{r_i(\beta)}{\hat{\sigma}_s}\right) + \lambda_n \sum_{j=1}^{p} |\beta_j|^{\gamma}. \tag{11}$$

The factor $\hat{\sigma}_s^2$ before the summation is added to ensure the resulting estimator coincides with the bridge regression estimator when $\rho_1(r) = r^2$. Setting the derivative of $L_n(\beta)$ with respect to $\beta$ to zero gives

$$\hat{\sigma}_s \sum_{i=1}^{n} \Psi_1\left(\frac{r_i(\beta)}{\hat{\sigma}_s}\right) \mathbf{x}_i - \lambda_n \gamma V(\beta) = \mathbf{0}, \tag{12}$$

where $V(\beta)$ is a $p \times 1$ vector of the form $(|\beta_1|^{\gamma-1} sgn(\beta_1), |\beta_2|^{\gamma-1} sgn(\beta_2), ..., |\beta_p|^{\gamma-1} sgn(\beta_p))^T$ and $sgn(.)$ is the signum function. Equation (12) can also be written as

$$(X^T W(\beta) X + \lambda_n \gamma W_0(\beta))\beta = X^T W(\beta) Y, \tag{13}$$

where $W(\beta) = diag(w_1(\beta), w_2(\beta), ..., w_n(\beta))$ and $w_i(\beta) = \Psi_1\left(\frac{r_i(\beta)}{\hat{\sigma}_s}\right) / \frac{r_i(\beta)}{\hat{\sigma}_s}$, for $i = 1, 2, 3, ..., n$

and $W_0(\beta) = diag\left(|\beta_1|^{\gamma-2}, |\beta_2|^{\gamma-2}, ..., |\beta_p|^{\gamma-2}\right)$ From (13) we get

$$\hat{\beta} = \left(X^T W(\hat{\beta}) X + \lambda_n \gamma W_0(\hat{\beta})\right)^{-1} X^T W(\hat{\beta}) Y, \tag{14}$$

provided that $\left(X^T W(\hat{\beta}) X + \lambda_n \gamma W_0(\hat{\beta})\right)$ is of full rank. Since, for an appropriate choice of $\rho_1$, the weight function $w(\mathbf{t}) = \Psi_1(t)/t$ is a decreasing function of $|t|$, data points with large residuals receive small weights, and hence will be downweighted.

Further, using $\hat{\beta}$ given in (14), $\hat{Y} = X\left(X^T W(\hat{\beta}) X + \lambda_n \gamma W_0(\hat{\beta})\right)^{-1} X^T W(\hat{\beta}) Y$, and the hat matrix $H = X\left(X^T W(\hat{\beta}) X + \lambda_n \gamma W_0(\hat{\beta})\right)^{-1} X^T W(\hat{\beta})$, respectively. The trace of $H$ can be used to choose the regularization parameters $\lambda_n$ and $\gamma$ with a *BIC* criterion.

When $\lambda_n = 0$, the solution of the minimization problem given in (10) is the MM estimator. When $\gamma = 2$ we obtain the robust ridge regression estimator proposed by Maronna (2011). The case $\gamma = 1$ yields the robust LASSO estimator based on the MM regression method.

## 2.3 Asymptotic Properties of MM-BR Estimator

We discuss the consistency, asymptotic normality and sparsity properties of the MM-BR estimator. Throughout, $\hat{\beta}_n$ denotes the minimizer of $L_n(\beta)$. That is, we will use $\hat{\beta}_n$ rather than $\hat{\beta}_{MM-BR}$ to simplify the notation. To study the asymptotic properties of the MM-BR estimator, we need the following additional assumptions.

A3: Let

$$l(\beta) = E_F\left(\rho_1\left(\frac{r(\beta)}{\sigma_0}\right)\right),$$

where $r(\beta) = y - \mathbf{x}^T \beta$ and $\sigma_0$ is defined by $E_F\left(\rho_0\left(\frac{r(\beta_0)}{\sigma_0}\right)\right) = 0.5$. Then, we assume that $l(\beta)$ has a unique minimum at $\beta_0$. The expectations being taken with respect to the error distribution, $F$.

Note that $\lim_{n\to\infty} \frac{1}{n}\sum_{i=1}^{n}\rho_1\left(\frac{r_i(\beta)}{\hat{\sigma}_s}\right) = E_F\left(\rho_1\left(\frac{r(\beta)}{\sigma_0}\right)\right) = l(\beta)$ in probability (see Yohai, 1985, Lemma 4.2. pp.38).

A4: $\rho_1(x)$ is twice continuously differentiable and there exists $m$ such that for $|x| \geq m$ $\rho_1(x) = a$.

A5: $\frac{1}{n}\sum_{i=1}^{n}\mathbf{x}_i\mathbf{x}_i^T \to C$ as $n \to \infty$ and $C$ is a nonsingular matrix.

*Consistency.* To explore the consistency of the MM-BR estimator we define

$$Z_n(\beta) = n^{-1}L_n(\beta) = \frac{\hat{\sigma}_s^2}{n}\sum_{i=1}^{n}\rho_1\left(\frac{r_i(\beta)}{\hat{\sigma}_s}\right) + \frac{\lambda_n}{n}\sum_{j=1}^{p}|\beta_j|^\gamma, \tag{15}$$

which is also minimized at $\beta = \arg\min_\beta L_n(\beta) = \hat{\beta}_n$. The following proposition shows that the MM-BR estimator is a consistent estimator of $\beta_0$, provided that $\lambda_n/n \to 0$ as $n \to \infty$.

**Proposition 1.** *Let* $\gamma > 0$ *and* $\lambda_n/n \to \lambda_0 \geq 0$ *as* $n \to \infty$. *Suppose that assumptions A1, A2, and A3 hold, and the initial estimator* $\hat{\beta}_s$ *is a consistent estimator for* $\beta_0$. *Then,* $\hat{\beta}_n \to \arg\min(Z(\beta))$ *in probability, where*

$$Z(\beta) = \sigma_0^2 l(\beta) + \lambda_0 \sum_{j=1}^{p}|\beta_j|^\gamma. \tag{16}$$

*Thus, if* $\lambda_n/n \to 0$ *as* $n \to \infty$, $\hat{\beta}_n \to \beta_0$ *in probability.*

In the proof of this proposition, $\lambda_n = o(n)$ is sufficient to have the consistency for the MM-BR estimator. But, in the proof of Proposition 2 the penalty regularization parameter should grow more slowly than the rate in the consistency proof to get the limiting distribution of the MM-BR estimator. If the penalty regularization parameter $\lambda_n$ grows too slowly, then the limiting distribution of $\hat{\beta}_n$ (the MM-BR estimator) will be the same as the limiting distribution of $\hat{\beta}_{MM}$. This was also discussed by Knight and Fu (2000) and Caner (2009) for the bridge regression estimator and the LASSO-type GMM estimator, respectively.

*Asymptotic normality.* The following proposition gives the limiting distribution of the MM-BR estimator for the cases $\gamma \geq 1$ and $\gamma < 1$.

**Proposition 2.** *Suppose that assumptions A1-A5 hold and the initial estimator* $\hat{\beta}_s$ *is a consistent estimator for* $\beta_0$. *Then,* $\hat{\beta}_n$ *satisfies the following.*
*(i) Let* $\gamma \geq 1$ *and* $\lambda_n/\sqrt{n} \to \lambda_0 \geq 0$ *as* $n \to \infty$. *Then,*

$$\sqrt{n}(\hat{\beta}_n - \beta_0) \to_d \arg\min_{\mathbf{u}} V(\mathbf{u}),$$

*where,* $\mathbf{u} = (u_1,...,u_p)^T$ *in* $R^p$, $\to_d$ *denotes convergence in distribution*,

$$V(\mathbf{u}) = -\mathbf{u}^T\mathbf{W} + \frac{1}{2}\mathbf{u}^T\left[B\left(\Psi_1\left(\frac{r(\beta_0)}{\sigma_0}\right)\right)C\right]\mathbf{u} + \lambda_0\sum_{j=1}^{p}u_j sgn(\beta_{oj})|\beta_{oj}|^{\gamma-1}$$

*if* $\gamma > 1$, *and*

$$V(\mathbf{u}) = -\mathbf{u}^T \mathbf{W} + \frac{1}{2}\mathbf{u}^T \left[ B\left(\Psi_1\left(\frac{r(\beta_0)}{\sigma_0}\right)\right) C \right] \mathbf{u} + \lambda_0 \sum_{j=1}^{p}[u_j sgn(\beta_{oj})I(\beta_{oj} \neq 0) + |u_j|I(\beta_{oj} = 0)]$$

*if* $\gamma = 1$. *The random vector,* $\mathbf{W}$, *has a* $N_p\left(\mathbf{0}, \sigma_0^2 A\left(\Psi_1\left(\frac{r(\beta_0)}{\sigma_0}\right)\right) C\right)$ *distribution,*

$$A\left(\Psi_1\left(\frac{r(\beta_0)}{\sigma_0}\right)\right) = E_F\left(\Psi_1^2\left(\frac{r(\beta_0)}{\sigma_0}\right)\right) \text{ and } B\left(\Psi_1\left(\frac{r(\beta_0)}{\sigma_0}\right)\right) = E_F\left(\Psi_1'\left(\frac{r(\beta_0)}{\sigma_0}\right)\right).$$

*(ii) Let* $\gamma < 1$ *and* $\lambda_n/n^{\gamma/2} \to \lambda_0 \geq 0$ *as* $n \to \infty$. *Then,*

$$\sqrt{n}(\hat{\beta}_n - \beta_0) \to_d \arg\min_{\mathbf{u}} V(\mathbf{u}),$$

*where*

$$V(\mathbf{u}) = -\mathbf{u}^T \mathbf{W} + \frac{1}{2}\mathbf{u}^T \left[ B\left(\Psi_1\left(\frac{r(\beta_0)}{\sigma_0}\right)\right) C \right] \mathbf{u} + \lambda_0 \sum_{j=1}^{p}|u_j|^{\gamma} I(\beta_{oj} = 0).$$

Note that when $\lambda_0 = 0$,

$$\sqrt{n}(\hat{\beta} - \beta_0) \to_d N\left(\mathbf{0}, \sigma_0^2 \frac{A\left(\Psi_1\left(\frac{r(\beta_0)}{\sigma_0}\right)\right)}{B\left(\Psi_1\left(\frac{r(\beta_0)}{\sigma_0}\right)\right)^2} C^{-1}\right), \tag{17}$$

which is the limiting distribution of the MM regression estimator given in Yohai (1987).

Further, when $\gamma \geq 2$ the distribution of $argminV(\mathbf{u})$ will be

$$\sqrt{n}(\hat{\beta}_n - \beta_0) \to_d N\left(-\left[B\left(\Psi_1\left(\frac{r(\beta_0)}{\sigma_0}\right)\right)C\right]^{-1} C^*(\beta_0)\beta_0, \sigma_0^2 \frac{A\left(\Psi_1\left(\frac{r(\beta_0)}{\sigma_0}\right)\right)}{B\left(\Psi_1\left(\frac{r(\beta_0)}{\sigma_0}\right)\right)^2} C^{-1}\right),$$

(18)

where, $C^*(\beta_0) = diag(|\beta_{01}|^{\gamma-2},...,|\beta_{0p}|^{\gamma-2})$. In particular, the limiting distribution of the robust ridge regression ($\gamma = 2$) based on the MM estimator (Maronna, 2011) is

$$\sqrt{n}(\hat{\beta}_n - \beta_0) \to_d N\left(-\left(B\left(\Psi_1\left(\frac{r(\beta_0)}{\sigma_0}\right)\right)C\right)^{-1}\beta_0, \sigma_0^2 \frac{A\left(\Psi_1\left(\frac{r(\beta_0)}{\sigma_0}\right)\right)}{B\left(\Psi_1\left(\frac{r(\beta_0)}{\sigma_0}\right)\right)^2}C^{-1}\right). \tag{19}$$

*Sparsity.* Suppose that the true parameter vector can be partitioned such that $\beta_0 = (\beta_{01}^T, \beta_{02}^T)^T$, where $\beta_{01}$ is a $p_0 \times 1$ and $\beta_{02}$ is a $(p-p_0) \times 1$ vector and that $\beta_{01} \neq \mathbf{0}$ and $\beta_{02} = \mathbf{0}$. Also, partition $\beta = (\beta_1^T, \beta_2^T)^T$ and $\hat{\beta}_n = (\hat{\beta}_{n1}^T, \hat{\beta}_{n2}^T)^T$ according to the partition of the true parameter vector. We do not know which coefficients are zero and which are nonzero. Further, partition $\mathbf{x}_i = (\mathbf{x}_{i1}^T, \mathbf{x}_{i2}^T)^T$, according to the partition of the true parameter vector. Then, we have the following proposition, which shows the sparsity property of the MM-BR estimator. The sparsity property means that small estimated coefficients are automatically set to zero by the estimator (e.g. see Fan and Li, 2001, pp.1349).

**Proposition 3.** *Assume that* $0 < \gamma \leq 1$. *If* $\lambda_n/n \to 0$ *and* $\lambda_n/\sqrt{n} \to \infty$ *as* $n \to \infty$. *Then the consistent MM-BR estimator* $\hat{\beta}_n = (\hat{\beta}_{n1}^T, \hat{\beta}_{n2}^T)^T$ *satisfies* $\hat{\beta}_{n2} = \mathbf{0}$ *with probability tending to 1.*

From the second part of Proposition 2, if $\gamma < 1$, then the nonzero regression parameters can be estimated at the standard rate without having asymptotic bias. On the other hand, the estimates of the zero regression parameters are shrunk to zero with positive probability. The first part of Proposition 2 shows that when $\gamma \geq 1$ the nonzero regression parameters are estimated with some asymptotic bias if $\lambda_0 > 0$. These results are analogous to the BR estimator results given by Knight and Fu (2000). Further, Proposition 3 shows that the estimator will have the sparsity property for the case $0 < \gamma \leq 1$. Thus, when $0 < \gamma < 1$ the proposed estimation method will provide variable selection for the MM regression and the estimator will be asymptotically unbiased. For $\gamma = 1$, the proposed estimation method, which is a robust LASSO based on MM estimation, will provide variable selection, but the resulting estimator will not be asymptotically unbiased. Finally, for the case $\gamma > 1$ the proposed estimation method shrinks the estimates of the regression parameters, but does not provide variable selection.

Proofs of these propositions are given in the Appendix.

## 3 Algorithm to Compute MM-BR Estimates

We propose an algorithm to minimize the penalized objective function given in (11). For the case $\gamma > 1$, the panelized objective function is everywhere differentiable with respect to $\beta$. Hence, it can be minimized using some standard gradient-based algorithms. However, for the case

$0 < \gamma \leq 1$, the panelized objective function given in (11) becomes non-differentiable at the origin as the penalty function is not differentiable at $\beta = 0$. This singularity problem makes it difficult to minimize the penalized objective function when $0 < \gamma \leq 1$. One way of handling a minimization problem like this is to approximate the penalty function into a convex function. This can be done either via the local quadratic approximations (LQA) proposed by Fan and Li (2001), or the local linear approximations (LLA) introduced by Zou and Li (2008). These approximations help to perform complicated optimization problems such as (10) with some standard algorithms. These two algorithms have been often used to find penalized least squares and penalized maximum likelihood estimates (Fan and Li, 2001; Hunter and Li, 2005; Liu et al. 2007; Zou and Li, 2008; Li and Yu, 2009; Li et al. 2011; Park and Yoon, 2011).

We use the LQA algorithm to compute the MM-BR estimates. We also give the least squares approximation estimator, and the one-step estimator based on the least squares approximation and the LQA algorithm.

### 3.1 LQA Algorithm

Suppose we assign an initial value $\beta^{(0)}$ that is close to the true value of $\beta$. If $\beta^{(0)}$ has some components that are very close to zero, then we set those components to zero (Fan and Li 2001; Zou and Li 2008; Park and Yoon 2011). We can take the unpenalized MM-estimate to be the initial value $\beta^{(0)}$. Fan and Li (2001) propose locally approximating to the penalty function by the following quadratic function

$$\left|\beta_j\right|^\gamma \approx \left|\beta_j^{(0)}\right|^\gamma + \frac{\gamma}{2}\left|\beta_j^{(0)}\right|^{\gamma-2}\left(\beta_j^2 - \left(\beta_j^{(0)}\right)^2\right)$$

at $\beta^{(0)} = (\beta_1^{(0)}, \beta_2^{(0)}, ..., \beta_p^{(0)})^T$. Using this approximation, the minimization problem (10) can be rewritten as

$$\hat{\beta}_{MM-BR} = \arg\min_{\beta} L_n^*(\beta), \qquad (20)$$

with the penalized objective function

$$L_n^*(\beta) = \frac{\hat{\sigma}_s^2}{n}\sum_{i=1}^n \rho_1\left(\frac{r_i(\beta)}{\hat{\sigma}_s}\right) + \frac{\lambda_n \gamma}{2n}\sum_{j=1}^p \left|\beta_j^{(0)}\right|^{\gamma-2}\beta_j^2$$

$$= \frac{\hat{\sigma}_s^2}{n}\sum_{i=1}^n \rho_1\left(\frac{r_i(\beta)}{\hat{\sigma}_s}\right) + \frac{\lambda_n \gamma}{2n}\beta^T W_0(\beta^{(0)})\beta.$$

$L_n^*(\beta)$ can be considered as the objective function for a robust weighted-ridge regression problem. In the robust ridge regression estimation proposed by Maronna (2011), the penalty function has the form $\sum_{j=1}^p \beta_j^2$, where we assign different weights to the different coefficients.

Setting the derivative of $L_n^*(\beta)$ with respect to $\beta$ to zero yields

$$\left(X^T W(\beta) X + \lambda_n \gamma W_0(\beta^{(0)})\right)\beta = X^T W(\beta) Y.$$

From this we get

$$\hat{\beta} = \left(X^T W(\hat{\beta}) X + \lambda_n \gamma W_0(\beta^{(0)})\right)^{-1} X^T W(\hat{\beta}) Y$$

provided that $\left(X^T W(\hat{\beta}) X + \lambda_n \gamma W_0(\beta^{(0)})\right)$ is of full rank. Given $\lambda_n$, $\gamma$ and $\beta^{(0)}$, the weighted estimator $\hat{\beta}$ given above suggests the following iteratively reweighting algorithm

$$\beta^{(k+1)} = \left(X^T W(\beta^{(k)}) X + \lambda_n \gamma W_0(\beta^{(k)})\right)^{-1} X^T W(\beta^{(k)}) Y, \tag{21}$$

for $k = 0,1,2,3,...$ The convergence point of the sequence $\{\beta^{(k)}\}$, for $k = 0,1,2,3,...$, can be taken as the MM-BR estimate for the regression parameter $\beta$.

When $\gamma < 2$ and some components of $\beta^{(k)}$ are very close to zero for some $j$ $|\beta_j^{(k)}| < c$ (a prespecified small cutoff value), there will be a numerical instability due to the diagonal elements of the $W_0(\beta^{(k)})$. To avoid this, Fan and Li (2001) suggest taking those components equal to zero and delete those components of **x** from the further iterations (see also Zou and Li, 2008). Also, Hunter and Li (2005) suggest replacing $|\beta_j^{(k)}|$ by $|\beta_j^{(k)} + c_0|$ to avoid the numerical instability, where $c_0$ is a prespecified small value of perturbation.

We should note that, similar to the LQA algorithm, we can use the LLA algorithm introduced by Zou and Li (2008) to perform minimization of the penalized objective function given in (11). However, since we use the LQA algorithm to compute the MM-BR estimates, we do not discuss the LLA algorithm in this paper.

### 3.2 Least Squares Approximation: One-Step Estimator

We use the least squares approximation to reduce the MM-bridge minimization problem (10) to a quadratic minimization problem. This approximation has been used by Fan and Li (2001), Wang and Leng (2007), and Zou and Li (2008) to solve penalized likelihood and LASSO problems. This approximation will allow us to define a one-step estimator based on LQA algorithm.

Using Taylor series expansion at $\hat{\beta}_{MM}$, the first term of (11) may be locally approximated by

$$\hat{\sigma}_s^2 \sum_{i=1}^n \rho_1\left(\frac{r_i(\beta)}{\hat{\sigma}_s}\right) \approx \hat{\sigma}_s^2 \sum_{i=1}^n \rho_1\left(\frac{r_i(\hat{\beta}_{MM})}{\hat{\sigma}_s}\right) \tag{22}$$

$$-\left[\hat{\sigma}_s \sum_{i=1}^n \Psi_1\left(\frac{r_i(\hat{\beta}_{MM})}{\hat{\sigma}_s}\right) \mathbf{x}_i\right]^T (\beta - \hat{\beta}_{MM})$$

$$+ \frac{1}{2}(\beta - \hat{\beta}_{MM})^T \left[ \sum_{i=1}^{n} \Psi_1' \left( \frac{r_i(\hat{\beta}_{MM})}{\hat{\sigma}_s} \right) \mathbf{x}_i \mathbf{x}_i^T \right] (\beta - \hat{\beta}_{MM})$$

Ignoring the constant term and noticing that $\sum_{i=1}^{n} \Psi_1 \left( \frac{r_i(\hat{\beta}_{MM})}{\hat{\sigma}_s} \right) \mathbf{x}_i = 0$, we have the following quadratic approximation

$$\hat{\sigma}_s^2 \sum_{i=1}^{n} \rho_1 \left( \frac{r_i(\beta)}{\hat{\sigma}_s} \right) \approx \frac{1}{2} (\beta - \hat{\beta}_{MM})^T (X^T W^*(\hat{\beta}_{MM}) X)(\beta - \hat{\beta}_{MM}), \tag{23}$$

where $W^*(\hat{\beta}_{MM}) = diag\left( \Psi_1' \left( \frac{r_1(\hat{\beta}_{MM})}{\hat{\sigma}_s} \right), ..., \Psi_1' \left( \frac{r_n(\hat{\beta}_{MM})}{\hat{\sigma}_s} \right) \right)$. Using this approximation, the penalized objective function given in (11) can be rewritten as

$$L_n(\beta) \approx \frac{1}{2} (\beta - \hat{\beta}_{MM})^T (X^T W^*(\hat{\beta}_{MM}) X)(\beta - \hat{\beta}_{MM}) \tag{24}$$
$$+ \lambda_n \sum_{j=1}^{p} |\beta_j|^{\gamma}.$$

The approximated objective function given above is the same as the bridge regression objective function given in (3). Therefore, with this approximation the original penalized MM-bridge regression minimization problem (10) can be rewritten as the asymptotically equivalent penalized least squares minimization problem

$$\tilde{\beta}_{MM-BR} = \arg \min_{\beta} \left[ \frac{1}{2} (\beta - \hat{\beta}_{MM})^T (X^T W^*(\hat{\beta}_{MM}) X)(\beta - \hat{\beta}_{MM}) + \lambda_n \sum_{j=1}^{p} |\beta_j|^{\gamma} \right]. \tag{25}$$

Moreover, if we use the local quadratic approximation (LQA) to the penalty function and set $\beta^{(0)} = \hat{\beta}_{MM}$, then we have the following one-step estimator

$$\tilde{\beta}_{MM-BR}^{(1)} = (D + \lambda_n \gamma W_0(\hat{\beta}_{MM}))^{-1} D \hat{\beta}_{MM}, \tag{26}$$

where $D = X^T W^*(\hat{\beta}_{MM}) X$. This is a ridge type estimator based on $\hat{\beta}_{MM}$.

## 4 Numerical Studies

### 4.1 Simulation

We discuss a simulation study to evaluate the finite-sample performance of the MM-BR

method in terms of variable selection and robust estimation of regression parameters for the case $n > p$. We compare the MM-BR method with the LASSO and the sparse LTS methods. The sparse LTS estimates are computed using the R package *robustHD* (robust methods for high dimensional data) (Alfons et al. 2013). The subset size, $h$, for the sparse LTS is taken as $\lfloor (n+1)0.75 \rfloor$, as suggested by Alfons et al. (2013). The R package *lars* is used to compute LASSO.

We use the iteratively reweighting algorithm given in (21) to compute the MM-BR estimates. We only consider the case $\gamma = 1$ to compare with the LASSO and sparse LTS. Note that $\gamma = 1$ corresponds to the robust LASSO based on MM regression method. Since $\gamma < 2$, we may encounter numerical instability problems. We use the approach suggested by Fan and Li (2001) to deal with the numerical instability problem. If $\beta_j^{(k)}$, the $j$th component of $\beta^{(k)}$, is very close to 0, $|\beta_j^{(k)}| < c$ (a predefined small cutoff value), then set $\beta_j^{(k)} = 0$ and delete the $j$th component of $\mathbf{x}$ from the iteration. The same approach was also used by Huang et al. (2008), Park and Yoon (2011), and Li et al. (2011) with different cutoff values. For example, Huang et al. (2008) and Park and Yoon (2011) take the cutoff values as $10^{-4}$ and $10^{-3}$, respectively. In our study we set $c = 10^{-5}$. We take the MM-estimate to be the initial value $\beta^{(0)}$ and stop the algorithm when $\max_{1 \le j \le p} \left| \beta_j^{(k+1)} - \beta_j^{(k)} \right| \le c.$

In the simulation study we use the Tukey bisquare $\rho$ function. The tuning parameter in $\rho$ and the penalty regularization parameter $\lambda_n$ can be chosen via a BIC-based tuning parameter selector proposed by Wang et al. (2007) (also see Li et al., 2011). That is, the optimal values of these parameters can be chosen by minimizing the criterion

$$BIC = n\log\left( \hat{\sigma}_s^2 \sum_{i=1}^{n} \rho_1\left( \frac{r_i(\hat{\beta}_{MM-BR})}{\hat{\sigma}_s} \right) \right) + trace(H)\log(n), \qquad (27)$$

where $H = X\left(X^T W(\hat{\beta}_{MM-BR})X + \lambda_n \gamma W_0(\hat{\beta}_{MM-BR})\right)^{-1} X^T W(\hat{\beta}_{MM-BR})$. To simplify the computation, we fix the tuning parameter of $\rho$ function to the value given in the Maronna (2011), and select only $\lambda_n$ using the $BIC$ criterion given above.

We perform several different simulations to assess the finite sample performance of the MM-BR estimator in terms of variable selection and robust estimation. For the variable selection, we report the average numbers of the correct and the incorrect zero coefficients in the final models. Concerning the modeling performance, we compute the model error, defined by $MSE = (\hat{\beta} - \beta_0)^T E(\mathbf{xx}^T)(\hat{\beta} - \beta_0).$

*Case 1*. In the first part of this case, we have the following data configuration. Let $\varepsilon$ be the contamination rate with values $0, 0.1$ and $0.2$; $n$ be the sample size with values $50$, $100$ and $200$; and $m = [\varepsilon n]$ be the number of contaminated data, where $[.]$ denotes the integer part. The data sets are generated as follows. We set $\beta_0 = (3, 1.5, 0, 0, 2, 0, 0, 0)^T$ and generate $n - m$ data from $\mathbf{x}_{1i} : N_p(\mathbf{0}, \mathbf{V})$ where $\mathbf{V} = (v_{ij})$ with $v_{ij} = 0.5^{|i-j|}$ and $p = 8$. The values of the response variable are generated according to the model $y_{1i} = \mathbf{x}_{1i}^T \beta_1 + \sigma \varepsilon_i$, for $i = 1, 2, ..., n - m$, where $\varepsilon$ is generated from the standard t-distributions with 1 (Cauchy distribution) and 3 degrees of freedom. These distributions allow us to have a heavy-tailed error distribution and some possible outliers in the $y$ direction. Two values of $\sigma$, 0.5 and 1, are taken. These $n-m$ data points will form the main bulk of the data. The $m$ contaminated data points are produced as follows. Generate $\mathbf{x}_{2i} : N_p(\mu, \mathbf{I})$ with $\mu \neq \mathbf{0}$ and take a vector $\beta_2 \neq \beta_1$. Then, use $y_{2i} = \mathbf{x}_{2i}^T \beta_2$, $i = 1, 2, ..., m$ to generate the values of the response variable of the contaminated observations. These $m$ points will form the contaminated part of the data set. We combine these two data sets to make one data set $(\mathbf{x}_i, y_i)$ for $i = 1, 2, ..., n$. Finally, we fit a linear regression model $y = \mathbf{x}^T \beta + \varepsilon$ to our data set, estimate the unknown parameter vector $\beta$ and select the significant variables.

In the second part of this simulation, the error terms follow a standard normal distribution and we apply two different contamination types. The same sample sizes given in the first part are considered. The first one is vertical outliers, which correspond to outliers in the response. In this case, $m = [\varepsilon n]$ of the error terms follow a normal distribution $N(25, \sigma^2)$. The second case is leverage points along with the vertical outliers. The leverage points are again generated using the leverage point generating strategy described in the first part. In both cases, the same contamination proportions given in the first part are applied.

*Case 2.* The second part of our simulation study is for the case $p = 50$. We again consider two different simulation configurations. For all the configurations we set $n = 100$, 300 and 500, and use the same values of $\varepsilon$ given in *Case 1*. The same data generating strategy described in *Case 1* is used to generate the predictors. In this case, the components of the coefficient vector $\beta_0$ are taken as $\beta_{0i} = 2$ for $1 \leq i \leq 10$ and $\beta_{0i} = 0$ for $11 \leq i \leq 50$. The response variable is generated according to the regression model $y_i = \mathbf{x}_i^T \beta_0 + \sigma \varepsilon_i$, for $i = 1, 2, ..., n$. Two values of $\sigma$, 0.5 and 1, are considered. In the first part of this case, the error term $\varepsilon$ is again generated from the Standard t-distributions with 1 (Cauchy distribution) and 3 degrees of freedom. In the second configuration of this case, the error terms follow a standard normal distribution. We again apply two different contamination structures, vertical outliers and leverage points. We use the same contamination proportions given in *Case 1*.

*Case 3.* We explore the performance of the estimators over a challenging leverage effect problem, which has been suggested by one of the referees, and I would like to thank them for this valuable contribution to the paper. Note that the same problem has also been considered by Alfons et al. (2013), and we use the same simulation plan from that work.

The good data points are generated using the same data generating plans described in *Case 1* and *Case 2*. The contaminated observations are generated as follows. First, we generate $m = [\varepsilon n]$ leverage points $\mathbf{x}_{2i}$, for $i = 1, 2, ..., m$, from $N_p(\mu_1, \mathbf{I})$ distribution, where $\mu_1 = (5, ..., 5)^T$. We generate values of the response variable corresponding to the leverage points using $y_{2i} = K \mathbf{x}_{2i}^T \beta_2$, for $i = 1, 2, ..., m$, with $\beta_2 = (-1/p, ..., -1/p)^T$, which is very different from the true parameter vectors given in the first two simulation settings. The scalar, $K$, which regulates the size of the contamination, takes values on a grid, searching for the worst behavior of each estimator. We take five different values for $K$ between 1 and 30. Finally, the values of $\varepsilon$ are taken as 0.1 and 0.2. For each $K$, we carry out 100 simulation runs, and then average the MSE values. Figure 3 displays the averaged MSEs of the estimates as a function of $K$.

In Tables 1-4, we summarize the simulation results for the LASSO, the sparse LTS and the MM-BR methods over the 100 simulated data sets. In the tables, column *Correct* shows the average number of zero coefficients correctly estimated to be zero, and column *Incorrect* shows the average number of nonzero coefficients incorrectly estimated to be zero. The columns *MeanMSE* and *MedianMSE* are the mean and the median of the MSE values over the simulated data sets.

Table 1 and Table 3 display the simulation results for *Case 1*. From Table 1, where there is no contamination the sparse LTS and MM-BR methods have similar variable selection performance with small mean and median MSE values. They are able to select the significant variables and do not incorrectly estimate the nonzero coefficients to be zero or zero coefficients to be nonzero. The performance of the LASSO is also comparable with the robust methods. When we introduce contamination in addition to the heavy-tailed error distributions, the LASSO, which is not robust, is noticeably influenced by the outliers with higher mean and median MSE values. On the other hand, the sparse LTS and the MM-BR methods both retain their good performance in terms of model selection and estimation.

In Table 3, we report the simulation results for the normally distributed error case. For the case $\sigma = 0.5$, the robust methods exhibit excellent performance for all the settings. LASSO also has similar performance for the case without contamination. If we introduce vertical outliers, the performance of robust methods are still good, but the performance of the LASSO worsens. The performance of the MM-BR appears better than the sparse LTS for $\sigma = 1$. In the case of leverage points, the robust methods still retain their excellent performance. However, the LASSO has very poor performance with very large MSE values.

In summary, the robust methods show excellent performance for all the simulation settings described in *Case 1*.

Figures 1, 2, and 3 show boxplots of the regression estimates $\hat{\beta}_1, ..., \hat{\beta}_8$ obtained using the MM, LASSO, sparse LTS and MM-BR methods. The results are based on 200 simulated data sets for the case $n = 100$, $\varepsilon = 0, 0.3$ and $\sigma = 0.5$, respectively. In all cases the error distribution is taken as the Cauchy distribution. For the sparse LTS we take $h = \lfloor (n+1)0.75 \rfloor$ for $\varepsilon = 0$ and $h = \lfloor (n+1)0.70 \rfloor, \lfloor (n+1)0.65 \rfloor$ for $\varepsilon = 0.3$. The horizontal dotted lines show the true parameter values.

LASSO is clearly influenced by the outliers. From Figure 1, even without contamination the estimates obtained from the LASSO tend to be more diffuse with noticeable biases. The robust methods have comparable performance in the first case. However, in the second setting, the MM-BR outperforms the sparse LTS in terms of selection and estimation. From Figure 2, the sparse LTS estimates have larger bias and dispersion relative to the MM-BR estimates. This shows that, while the sparse LTS with 0.3 trimming proportion does not yet break down for 0.3 contamination, it may have substantial bias. On the other hand, if the trimming proportion is chosen larger than 0.3, the performance of the sparse LTS becomes comparable with the MM-BR (Figure 2).

Tables 2 and 4 contain the simulation results for *Case 2*. From Table 2, in settings without contamination and heavy-tailed error distributions, the MM-BR method performs the best in terms of finding zero coefficients. LASSO and the sparse LTS methods have comparable behavior with the MM-BR method in terms of finding the zero coefficients. However, LASSO also has a problem of incorrectly estimating the nonzero coefficients to be zero.

When contamination is introduced, the MM-BR method performs slightly better than the sparse LTS method in terms of identifying zero coefficients. Neither of the robust methods incorrectly estimate nonzero coefficients to be zero. The mean and median MSE values are larger when the error distribution is Cauchy. In this case, the performance of the LASSO degrades with very high MSE values. It is not able to estimate the zero and the nonzero coefficients correctly.

The results of the second simulation configuration given in *Case 2* are presented in Table 4. Without contamination the robust methods again have excellent performance. The performance of the LASSO is inferior to the robust methods, but it still has reasonable behavior in terms of model selection.

In the case of vertical outliers, the robust methods MM-BR and the sparse LTS retain their excellent performance. the LASSO performance is not comparable with the performance of robust methods, but it still exhibits reasonable performance in terms of estimating the zero coefficients correctly. However, it has the same problem of identifying the nonzero coefficients to be zero.

When leverage points are introduced, the robust methods again exhibit excellent performance with small MSE values. On the other hand, the LASSO suffers significantly, which is reflected in very high MSE values and very poor model selection performance. In both simulation settings of *Case 2*, MM-BR and sparse LTS have larger MSE values for the case $\sigma = 1$.

The simulation results for *Case 3* are shown in Figure 4. The sparse LTS and MM-BR have similar performance for the case $\varepsilon = 0.1$ and $p = 8$. The MSE values increase for some intermediate values of $K$, and then decrease. This shows that the performance of the estimators are worsening for some intermediate size of the leverage effect, then their performance starts improving in terms of MSE. For larger values of $K$, the MM-BR has slightly smaller MSE values than the sparse LTS (first row of Figure 4). As expected, for both contamination rates the MSE for the LASSO is significantly increasing with increasing values of $K$. When $\varepsilon = 0.2$ and $p = 8$ (second row of Figure 4), the MSEs for the MM and MM-BR are lower than the MSE of the sparse LTS, and the MSE graph for the MM-BR has similar shape to the first case. On the other hand, the MSE plot for the sparse LTS increases at the beginning, and then decreases with increasing $K$. Alfons et al. (2013) reported the same behavior for the root mean squared prediction error of the sparse LTS for some of their simulation schemes. For the case $p = 50$ and $epsilon = 0.1$, while the MSE of the sparse LTS increases, that of MM-BR decreases with increasing $K$ (similar behavior is observed for the case $p = 50$ and $epsilon = 0.1$). We notice that for all the cases explored in this simulation setting, the MSE plot for the MM-BR estimates shows decreasing with increasing $K$. Thus, MM-BR is not influenced by the leverage points. Overall, we conclude that among these three penalized regression methods, the MM-BR method has the best performance against the leverage points.

Table 5 reports some simulation results for the one step MM-BR estimator given in (26). The overall performance of the one step version is inferior to the MM-BR estimator in terms of finding the zero coefficients. It also shows the problem of incorrectly estimating nonzero coefficients to be zero. The performance of the one step estimator is enhanced with increasing sample size.

Our simulation study is limited and we only considered the cases $n > p$ and $\gamma = 1$. However, we can say that the overall performance of the MM-BR method is comparable with the sparse LTS method in terms of model selection and robust estimation. The one step MM-BR estimator performance is inferior to the MM-BR estimator, but outperforms the LASSO. Thus our simulation study confirms that LASSO model selection methods are not robust to outliers in the data, and therefore robust model selection methods are needed.

## 4.2 Example

We use a pollution data set to evaluate the performance of the MM-BR method on real data. The pollution data, which can be obtained from the *SMPracticals* package in R, has been previously analyzed by McDonald and Schwing (1973), Luo et al.(2006), Park and Yoon (2011), and Kawano (2013), and is used here to evaluate the performance of the model selection methods.

The data set consists of 60 observations and 15 covariates. The response variable is the total-age adjusted mortality rates in 60 Standard Metropolitan Statistical Areas of the USA obtained for the years 1959–1961. The data are on the relationships between weather, socioeconomic, and air pollution variables and mortality rates.

After some preliminary analysis of the explanatory variables we observe that the data set may contain some leverage points. Observations 29, 48,47,49,18 and 32 have larger robust Mahalanobis distances, so these points may be potential leverage points. Therefore, using robust model selection methods may provide better results for this data set.

Table 6 shows the variable selection results using the whole data set obtained from the LASSO, sparse LTS, bridge estimate based on the OLS method for $\gamma = 0.7$ and MM-BR estimates for the cases $\gamma = 1$ and $0.7$. All the methods chose the variables $X_1, X_2, X_3, X_8$, and $X_{14}$. The variables $X_1, X_2$, and $X_3$ are weather conditions, $X_8$ is related to the population in urbanized areas, and $X_{14}$ is associated with the relative sulphur dioxide pollution. These variables are highly significant for the response variable. The MM-BR and sparse LTS procedures tended to choose more variables than the non-robust methods. The variables $X_{15}$ and $X_{11}$, related to weather condition and socioeconomic status, respectively, were selected by the robust methods but not by the non-robust methods. Also, $X_9$, which is the percentage of non-white population in urbanized areas, was selected by the MM-BR method and the non-robust methods, but sparse LTS did not select it. On the other hand, $X_6$, which is the median school years completed by those over 22, was selected by the sparse LTS and the non-robust methods, but not by the MM-BR method. Variables $X_4, X_{10}$, and $X_{12}$ were not included in any of the selected models, suggesting they are not significant for the response variable.

To validate the prediction errors, we randomly selected 40 data points for model fitting. The remaining 20 observations were used as the test data set. To avoid any bias related to a particular division of the data, we repeated this procedure 10 times. Table 7 summarizes the mean prediction errors for the LASSO, sparse LTS, bridge estimate with $\gamma = 0.7$, and MM-BR estimates for the cases $\gamma = 1$ and $0.7$. The MM-BR with $\gamma = 0.7$ has the smallest prediction error among all the methods. It also has the smallest model size among the robust methods.

## 5 Conclusions and discussion

We proposed the MM-BR method to improve the robustness of the LS based bridge regression method. The MM-BR method combines the MM and bridge regression methods to provide a robust bridge regression method. We explored the asymptotic properties of the MM-BR estimators, and showed that consistency, asymptotic normality and sparsity hold, under appropriate regularity conditions.

We provided an algorithm to compute MM-BR estimates, and have performed a simulation study and real-data example to illustrate the performance of the MM-BR estimator in terms of robust estimation and variable selection. Our limited simulation study confirmed that the MM-BR method (MM-LASSO) behaves comparably well in terms of variable selection and retains appealing robustness property of the MM regression method.

We have only considered the case $n > p$, but we believe that the results can be extended to the case $p > n$. We take the MM regression estimator as the initial solution for the MM-BR estimation problem, but if the number of variables exceeds the number of observations we cannot use the MM estimator as the initial solution, and our method will not be applicable. To overcome this problem we can apply the procedure proposed by Fan and Lv (2008) and Li et al. (2011). They suggest a two-stage procedure. First, a dimension reduction method to reduce the data dimensionality to smaller or equal to the sample size. Then a model selection method used to estimate the regression parameters and select the significant variables. Therefore, for high dimensionality problems, we can first use dimension reduction methods, and then apply the MM-BR method to perform the variable selection and the robust estimation simultaneously. This problem deserves further study and will be our next concern.

## Appendix

The technical proofs of the propositions given in Section 2 are included in this section.

*Proof of Proposition 1.* The proof of Proposition 1 will be similar to the consistency proofs given in Knight and Fu (2000). Caner (2009) also gives a similar proof to show the consistency of the LASSO-type GMM (generalized method of moments) estimator.

Note that if assumptions A1, A2 and A3 are satisfied and, if $\hat{\beta}_s$ is consistent for $\beta_0$, then $\hat{\sigma}_s$ is consistent estimator for $\sigma_0$ (Yohai, 1985, 1987).

Define $Z_n(\beta)$ as in (15) (Section 2). To prove the consistency of $\hat{\beta}_n$ we first need to show that

$$\sup_{\beta \in \Omega} |Z_n(\beta) - Z(\beta)| \to 0,$$

in probability (we have to show that $Z_n(\beta)$ converges uniformly in probability to $Z(\beta)$ defined in equation (16), (e.g., see Newey and McFadden, 1994, Theorem 2.1, pp.2121; Van der Vaart and Wellner, 1996, Theorem 3.2.2 and Corollary 3.2.3, pp.286-287; Van der Vaart, 2000, Theorem 5.7, pp.45; Knight and Fu, 2000, Theorem 1). Also, we have to show that $\hat{\beta}_n = O_p(1)$ ($\hat{\beta}_n$ is uniformly bounded in probability).

We will first show that $\frac{1}{n}\sum_{i=1}^{n}\rho_1\left(\frac{r_i(\beta)}{\hat{\sigma}_s}\right)$ converges uniformly in probability to $l(\beta)$ given in A3. This will be sufficient to show that $Z_n(\beta)$ converges uniformly in probability to $Z(\beta)$, because the second term in $Z_n(\beta)$ is not stochastic and the parameter space is compact (by assumption). We can use Lemma 2.4 (pp.2129) in Newey and McFadden (1994) to show the uniform convergence of $\frac{1}{n}\sum_{i=1}^{n}\rho_1\left(\frac{r_i(\beta)}{\hat{\sigma}_s}\right)$ to $l(\beta)$. We already have the compactness of the parameter space (by assumption) and the continuity of $\rho_1$ (assumption A1). Further, since

$\sup \rho_1 < \infty$ (assumption A1), $\rho_1\left(\dfrac{r(\beta)}{\sigma_0}\right) \le \sup_{\beta \in \Omega} \rho_1\left(\dfrac{r(\beta)}{\sigma_0}\right)$ and $E\left(\sup_{\beta \in \Omega} \rho_1\left(\dfrac{r(\beta)}{\sigma_0}\right)\right) < \infty$, we also have the other conditions stated in that Lemma. Then, under these conditions, we have that $E_F\left(\rho_1\left(\dfrac{r(\beta)}{\sigma_0}\right)\right)$ is continuous and

$$\sup_{\beta \in \Omega}\left|\frac{1}{n}\sum_{i=1}^{n}\rho_1\left(\frac{r_i(\beta)}{\hat{\sigma}_s}\right) - l(\beta)\right| \to 0$$

in probability. Finally, if we use this result, consistency of $\hat{\sigma}_s$ for $\sigma_0$, and $\lambda_n/n \to \lambda_0 \ge 0$ as $n \to \infty$, we get that

$$\sup_{\beta \in \Omega}|Z_n(\beta) - Z(\beta)| \to 0$$

in probability. (One can see the proof of Theorem 1 in Caner (2009) for a similar discussion). Moreover, since $Z_n(\beta) \ge \dfrac{\hat{\sigma}_s^2}{n}\sum_{i=1}^{n}\rho_1\left(\dfrac{r_i(\beta)}{\hat{\sigma}_s}\right)$ for all $\beta$, and since

$$\arg\min_{\beta}\left(\frac{\hat{\sigma}_s^2}{n}\sum_{i=1}^{n}\rho_1\left(\frac{r_i(\beta)}{\hat{\sigma}_s}\right)\right) = \hat{\beta}_{MM} = O_p(1)$$

(see Yohai, 1985, Lemma 4.1, pp.36-37), it follows that

$$\arg\min_{\beta}\left(Z_n(\beta)\right) = \hat{\beta}_n = O_p(1).$$

So, combining these two results we obtain that

$$\hat{\beta}_n \to \arg\min_{\beta} Z(\beta)$$

in probability. Further, when $\lambda_n/n \to 0$ as $n \to \infty$, $Z_n(\beta)$ converges uniformly in probability to $\sigma_0^2\ l(\beta)$ and, since $l(\beta)$ has a unique minimum at $\beta_0$ (assumtion A3) we get the consistency result: $\hat{\beta}_n \to \beta_0$ in probability.

*Proof of Proposition 2.* (i) To obtain the limiting distribution of $\hat{\beta}_n$, we will first define the following localized criterion function $V_n(\mathbf{u})$, where $\mathbf{u} = (u_1,...,u_p)^T$ is a local parameter vector in $R^p$ (e.g., see, Van der Vaart and Wellner, 1996, pp.279-88; Knight and Fu, 2000; Caner, 2009).

$$V_n(\mathbf{u}) = \hat{\sigma}_s^2 \sum_{i=1}^{n}\left(\rho_1\left(\frac{r_i(\beta_0 + \mathbf{u}/\sqrt{n})}{\hat{\sigma}_s}\right) - \rho_1\left(\frac{r_i(\beta_0)}{\hat{\sigma}_s}\right)\right) + \lambda_n \sum_{j=1}^{p}\left(|\beta_{oj} + u_j/\sqrt{n}|^\gamma - |\beta_{oj}|^\gamma\right).$$

Note that $V_n(\mathbf{u})$ is minimized at $\hat{\mathbf{u}}_n = \sqrt{n}(\hat{\beta}_n - \beta_0)$. To obtain the asymptotic distribution of $\hat{\beta}_n$ first we have to show that

$$V_n(\mathbf{u}) \to_d V(\mathbf{u}).$$

Using Taylor series expansion around $\mathbf{u} = \mathbf{0}$, we get

$$\hat{\sigma}_s^2 \sum_{i=1}^{n}\left(\rho_1\left(\frac{r_i(\beta_0 + \mathbf{u}/\sqrt{n})}{\hat{\sigma}_s}\right) - \rho_1\left(\frac{r_i(\beta_0)}{\hat{\sigma}_s}\right)\right) = -\mathbf{u}^T\left[\frac{\hat{\sigma}_s}{\sqrt{n}}\sum_{i=1}^{n}\Psi_1\left(\frac{r_i(\beta_0)}{\hat{\sigma}_s}\right)\mathbf{x}_i\right]$$
$$+ \frac{1}{2}\mathbf{u}^T\left[\frac{1}{n}\sum_{i=1}^{n}\Psi_1'\left(\frac{r_i(\beta_0)}{\hat{\sigma}_s}\right)\mathbf{x}_i\mathbf{x}_i^T\right]\mathbf{u}.$$

Since, $\dfrac{1}{n}\sum_{i=1}^{n}\Psi_1'\left(\dfrac{r_i(\beta_0)}{\hat{\sigma}_s}\right)\mathbf{x}_i\mathbf{x}_i^T \to B\left(\Psi_1'\left(\dfrac{r(\beta_0)}{\sigma_0}\right)\right)C$ and $\dfrac{\hat{\sigma}_s}{\sqrt{n}}\sum_{i=1}^{n}\Psi_1\left(\dfrac{r_i(\beta_0)}{\hat{\sigma}_s}\right)\mathbf{x}_i \to_d \mathbf{W}$ with

$\mathbf{W}: N_p\left(\mathbf{0}, \sigma_0^2 A\left(\Psi_1\left(\dfrac{r(\beta_0)}{\sigma_0}\right)\right)C\right)$ (Yohai, 1985), we get that

$$\hat{\sigma}_s^2 \sum_{i=1}^{n}\left(\rho_1\left(\frac{r_i(\beta_0 + \mathbf{u}/\sqrt{n})}{\hat{\sigma}_s}\right) - \rho_1\left(\frac{r_i(\beta_0)}{\hat{\sigma}_s}\right)\right) \to_d -\mathbf{u}^T\mathbf{W} + \frac{1}{2}\mathbf{u}^T\left[B\left(\Psi_1'\left(\frac{r(\beta_0)}{\sigma_0}\right)\right)C\right]\mathbf{u}.$$

Concerning the second part of $V_n(\mathbf{u})$, as in the proof of Theorem 2 in Knight and Fu (2000), if $\gamma > 1$ then

$$\lambda_n \sum_{j=1}^{p}\left(|\beta_{oj} + u_j/\sqrt{n}|^\gamma - |\beta_{oj}|^\gamma\right) = \frac{\lambda_n \gamma}{\sqrt{n}} \sum_{j=1}^{p} u_j sgn(\beta_{oj})|\beta_{oj}|^{\gamma-1}$$
$$\to \lambda_0 \gamma \sum_{j=1}^{p} u_j sgn(\beta_{oj})|\beta_{oj}|^{\gamma-1},$$

as $n \to \infty$, and if $\gamma = 1$, we have

$$\lambda_n \sum_{j=1}^{p}\left(|\beta_{oj} + u_j/\sqrt{n}|^\gamma - |\beta_{oj}|^\gamma\right) = \frac{\lambda_n}{\sqrt{n}} \sum_{j=1}^{p}[u_j sgn(\beta_{oj})I(\beta_{oj} \neq 0) + |u_j|I(\beta_{oj} = 0)]$$

$$\to \lambda_0 \sum_{j=1}^{p} [u_j sgn(\beta_{oj}) I(\beta_{oj} \neq 0) + |u_j| I(\beta_{oj} = 0)]$$

as $n \to \infty$. Therefore, combining these results, we get

$$V_n(\mathbf{u}) \to_d V(\mathbf{u}) \text{ for } \gamma \geq 1.$$

Further, $V_n(\mathbf{u})$ can be written as

$$V_n(\mathbf{u}) = -\mathbf{u}^T \left[ \frac{\hat{\sigma}_s}{\sqrt{n}} \sum_{i=1}^{n} \Psi_1\left(\frac{r_i(\beta_0)}{\hat{\sigma}_s}\right) \mathbf{x}_i \right] + \frac{1}{2} \mathbf{u}^T \left[ \frac{1}{n} \sum_{i=1}^{n} \Psi_1'\left(\frac{r_i(\beta_0)}{\hat{\sigma}_s}\right) \mathbf{x}_i \mathbf{x}_i^T \right] \mathbf{u}$$
$$+ \lambda_n \sum_{j=1}^{p} \left( \left|\beta_{oj} + u_j/\sqrt{n}\right|^\gamma - \left|\beta_{oj}\right|^\gamma \right)$$

Thus, since $V(\mathbf{u})$ has a unique minimum and $V_n(\mathbf{u})$ can be approximated by a convex function, it follows that (Geyer, 1996; Knight and Fu, 2000)

$$\arg\min_{\mathbf{u}} V_n(\mathbf{u}) = \sqrt{n}(\hat{\beta}_n - \beta_0) \to_d \arg\min_{\mathbf{u}} V(\mathbf{u}).$$

(ii) The proof of this part will be similar to the first part with some additional care due to the nonconvexity of the penalty function. As in the proofs of Theorem 3 in Knight and Fu (2000) and Theorem 2 in Caner (2009), if $\gamma < 1$ and $\frac{\lambda_n}{n^{\gamma/2}} \to \lambda_0 \geq 0$, we get

$$\lambda_n \sum_{j=1}^{p} \left( \left|\beta_{oj} + u_j/\sqrt{n}\right|^\gamma - \left|\beta_{oj}\right|^\gamma \right) = \frac{\lambda_n}{n^{\gamma/2}} \sum_{j=1}^{p} |u_j|^\gamma I(\beta_{oj} = 0)$$
$$\to \lambda_0 \sum_{j=1}^{p} |u_j|^\gamma I(\beta_{oj} = 0)$$

and the convergence is uniform over $\mathbf{u}$ in compact sets. Combining with the result obtained in the first part, it follows that

$$V_n(\mathbf{u}) \to_d V(\mathbf{u}).$$

To prove that $\arg\min_{\mathbf{u}} V_n(\mathbf{u}) = \sqrt{n}(\hat{\beta}_n - \beta_0) \to_d \arg\min_{\mathbf{u}} V(\mathbf{u})$ we have to show that $\arg\min_{\mathbf{u}} V_n(\mathbf{u}) = O_p(1)$ ($\arg\min_{\mathbf{u}} V_n(\mathbf{u})$ is uniformly bounded in probability). Note that, for all $\mathbf{u}$ and $n$ sufficiently large we can have

$$V_n(\mathbf{u}) \geq -\mathbf{u}^T \left[ \frac{\hat{\sigma}_s}{\sqrt{n}} \sum_{i=1}^{n} \Psi_1\left(\frac{r_i(\beta_0)}{\hat{\sigma}_s}\right) \mathbf{x}_i \right] + \frac{1}{2} \mathbf{u}^T \left[ \frac{1}{n} \sum_{i=1}^{n} \Psi_1'\left(\frac{r_i(\beta_0)}{\hat{\sigma}_s}\right) \mathbf{x}_i \mathbf{x}_i^T \right] \mathbf{u} - \lambda_n \sum_{j=1}^{p} \left( \left|u_j/\sqrt{n}\right|^\gamma \right)$$

$$\geq -\mathbf{u}^T\left[\frac{\hat{\sigma}_s}{\sqrt{n}}\sum_{i=1}^{n}\Psi_1\left(\frac{r_i(\beta_0)}{\hat{\sigma}_s}\right)\mathbf{x}_i\right] + \frac{1}{2}\mathbf{u}^T\left[\frac{1}{n}\sum_{i=1}^{n}\Psi_1'\left(\frac{r_i(\beta_0)}{\hat{\sigma}_s}\right)\mathbf{x}_i\mathbf{x}_i^T\right]\mathbf{u} - (\lambda_0+\delta)\sum_{j=1}^{p}\left|u_j\right|^{\gamma}$$
$$= V_n^l(\mathbf{u}),$$

where $\delta$ is a constat chosen to ansur $\frac{\lambda_n}{n^{\gamma/2}} \leq (\lambda_0+\delta)$ (see the proofs of Theorem 3 in Knight and Fu, 2000; and Theorem 2 in Caner, 2009). Since the second term (the quadratic term in $\mathbf{u}$) of $V_n^l(\mathbf{u})$ grows faster than the $\left|u_j\right|^{\gamma}$ terms (the quadratic term dominates the term with $\left|u_j\right|^{\gamma}$), we get $\arg\min_{\mathbf{u}} V_n^l(\mathbf{u}) = O_p(1)$. So, by the inequality $V_n(\mathbf{u}) \geq V_n^l(\mathbf{u})$, it follows that $\arg\min_{\mathbf{u}} V_n(\mathbf{u}) = O_p(1)$. Since $\arg\min_{\mathbf{u}} V(\mathbf{u})$ is unique with probability 1 (see Knight and Fu, 2000) we obtain that $\arg\min_{\mathbf{u}} V_n(\mathbf{u}) = \sqrt{n}(\hat{\beta}_n - \beta_0) \to_d \arg\min_{\mathbf{u}} V(\mathbf{u})$.

*Proof of Proposition 3.* The claim given in Proposotin 3 can also be rephrased as follows (see Fan and Li, 2001). If $\lambda_n/n \to 0$ and $\lambda_n/\sqrt{n} \to \infty$ as $n \to \infty$, then with probability tending to 1, for any given $\beta_1$ around $\beta_{01}$ and any constant $a$,

$$Z_n(\beta_1, \mathbf{0}) = \min_{\|\beta_2\| \leq a/\sqrt{n}} Z_n(\beta_1, \beta_2).$$

To prove this, it is sufficient to show that for some small $e_n = a/\sqrt{n}$ and $j = p_0+1,...,p$,

$$\frac{\partial Z_n(\beta)}{\partial \beta_j} < 0 \text{ for } -e_n < \beta_j < 0$$
$$\frac{\partial Z_n(\beta)}{\partial \beta_j} > 0 \text{ for } \quad 0 < \beta_j < e_n. \tag{28}$$

Consider the first derivative of $L_n(\beta)$ at any differentiable point $\beta = (\beta_1, \beta_2,...,\beta_p)^T$ with respect to $\beta_j$, $j = j = p_0+1,...,p$,

$$\frac{\partial Z_n(\beta)}{\partial \beta_j} = -(\hat{\sigma}_s/n)\sum_{i=1}^{n}\Psi_1\left(\frac{r_i(\beta)}{\hat{\sigma}_s}\right)x_{ij} + (\lambda_n/n)\gamma sgn(\beta_j)\left|\beta_j\right|^{\gamma-1}.$$

Since $\left[\frac{\hat{\sigma}_s}{\sqrt{n}}\sum_{i=1}^{n}\Psi_1\left(\frac{r_i(\beta)}{\hat{\sigma}_s}\right)\mathbf{x}_i\right] \to_d Y$ and the distribution of $Y$ is $N\left(\mathbf{0}, \sigma_0^2 A\left(\Psi_1\left(\frac{r(\beta_0)}{\sigma_0}\right)\right)C\right)$ we have $\frac{\hat{\sigma}_s}{\sqrt{n}}\sum_{i=1}^{n}\Psi_1\left(\frac{r_i(\beta)}{\hat{\sigma}_s}\right)\mathbf{x}_i = O_p(1)$ and $\frac{\hat{\sigma}_s}{\sqrt{n}}\sum_{i=1}^{n}\Psi_1\left(\frac{r_i(\beta)}{\hat{\sigma}_s}\right)\mathbf{x}_i$ is bounded in probability. These are due to the facts given by Shao (1999, pp.42). Thus, we get

$$\frac{\partial Z_n(\beta)}{\partial \beta_j} = (\lambda_n/n)\left\{-O_p\left(\frac{\sqrt{n}}{\lambda_n}\right) + \gamma sgn(\beta_j)|\beta_j|^{\gamma-1}\right\}.$$

If $\lambda_n/n \to 0$ and $\lambda_n/\sqrt{n} \to \infty$ as $n \to \infty$ the sign of the derivative is determined by the sign of $\beta_j$. This shows that the equations (1)-(2) hold, and hence the proof is completed.

Note that the methodology applied to prove Lemma 2 in the paper by Huang et al. (2008) can also be used to carry out the proof of this proposition.

# References

Alfons, A., Croux, C., and Gelper, S. (2013). Sparse least trimmed squares regression for analyzing high-dimensional large data sets, *The Annals of Applied Statistics, Vol.7, pp.226-248.*

Alfons, A. (2012). *robustHD:* Robust methods for high-dimensional data. R package version 0.1.0.

Antoniadis, A. and Fan, J. (2001). Regularization of Wavelets Approximation, *Journal of the American Statistical Association, Vol.96, pp.936-967.*

Arslan, O. (2012). Weighted LAD-LASSO method for robust parameter estimation and variable selection. *Computational Statistics and Data Analysis, Vol.56, pp.1952-1965.*

Armagan, A. (2009). Variational bridge regression, *Proceedings of the 12th International Conference on Artificial Intelligence and Statistics (AISTATS), Volume 5of JMLR:W&CP 5. Clearwater Beach Florida, USA.*

Caner, M. (2009). LASSO-type GMM estimator, *Econometric Theory, Vol.25, pp.270-290.*

Fan, J., and Li, R. (2001). Variable selection via nonconcave penalized likelihood and its oracle properties, *Journal of the American Statistical Association, Vol. 96, pp.1348-1360.*

Fan, J., and Lv, J. (2008). Sure independence screening for ultra-high dimensional feature space (with discussion), *Journal of the Royal Statistical Society B, Vol. 70, pp. 849-911.*

Frank, I.E., and Friedman, J. (1993). A statistical view of some chemometrics regression tools. *Thechnometrics, Vol. 35, pp. 109-148.*

Fu, W. J. (1998). Penalized regression: The bridge versus lasso. *Journal of Computational and Graphical Statistics, Vol.7, pp.397-416.*

Geyer, C. J. (1996). On the asymptotics of convex stochastic optimization. Unpublished manuscript

Huang, J., Horowitz, J.L., and Ma, S. (2008). Asymptotic properties of bridge estimators in sparse high-dimensional regression models. *The Annals of Statistics, 30,pp.587-613.*

Hunter, D.R., and Li, R. (2005). Variable selection using MM algorithms. *The Annals of Statistics, Vol.33, pp.1617-1642.*

Kawano, S. (2013). Selection of tuning parameters in bridge regression models via Bayesian information criterion, *Statistical Papers*, DOI 10.1007/s00362-013-0561-7.

Knight, K., and Fu, W. (2000). Asymptotics for lasso-type estimators. *The Annals of Statistics, Vol.28, pp.1356-1378.*

Li, B., and Yu, Q. (2009). Robust and sparse bridge regression. *Statistics and Its Interface, Vol.2, pp.481-491.*

Li, G., Peng, H., and Zhu, L. (2011). Nonconcave penalized M-estimation with a diverging number of parameters. *Statistica Sinica, Vol. 21, pp. 391-419.*

Liu, Y., Zhang, H.H., Park, C., and Ahn, J. (2007). Support vector machines with adaptive $L_q$ penalty. *Computational Statistics & Data Analysis, Vol. 51, pp.6380-6394.*

Luo, X., Stefanski,L. and Boos, D. D.(2006) Tuning variable selection procedures by adding noise, *Technometrics, Vol. 48. pp. 165-175.*

Maronna, R.A., Martin, R.D. and Yohai, V.J. (2006). *Robust Statistics Theory and Methods*, John Wiley and Sons, New York.

Maronna, R.A. and Yohai, V.J. (2010). Correcting MM estimates for fat data sets, *Computational Statistics & Data Analysis, Vol. 54, pp.3168-3173.*

Maronna, R.A. (2011). Robust ridge regression for high-dimensional data. *Technometrics, Vol. 53, pp.44-53.*

McDonald, G. and Schwing, R. (1973). Instabilities of regression estimates relating air pollution to mortality, *Technometrics, Vol. 15. pp. 463-482.*

Newey, W.K. and McFadden, D. (1994). Large sample estimation and hypothesis testing. In R.F. Engle and D.L. McFadden (eds.), *Handbook of Econometrics, vol. 4, pp. 2111-2245.*

Owen, A.B.,(2007). A robust hybrid of lasso and ridge regression. *Contemparary Mathematics, Vol. 443, pp.59-71.*

Park, C., and Yoon, Y. J. (2011). Bridge regression: Adaptivity and group selection. *Journal of Statistical Planning and Inference, Vol. 141, pp. 3506-3519.*

Wang, H., and Leng, C. (2007). Unified LASSO estimation by least squares approximation. *Journal of the American Statistical Association, Vol. 102, pp. 1039-1048.*

Wang, H., Li, R. Tsai, C.L. (2007). Tuning parameter selectors fort he smoothly clipped absolutle deviation method. textitBiometrika, 94, 553-568.

Xu, J. and Ying, Z., (2010). Simultaneous estimation and variable selection in median regression using Lasso-type penalty. *Annals of the Institute of Statistical Mathematics,* 62, 487-514.

Tibshiran, R. (1996). Regression shrinkage and selection via the lasso. *Journal of the Royal Statistical Society B, Vol. 58, pp. 267-288.*

Van der Vaart, A.W. and Wellner, J. A. (1996). *Weak Convergence and Empirical Processes.* Springer-Verlag.

Van der Vaart, A.W. (2000). *Asymptotic Statistics.* Cambridge Series in Statistical and Probabilistic Mathematics, Cambridge University Press.

Yohai, V.J. (1985). High breakdown point and high efficiency robust estimates for regression. *Technical Report No.66, Department of Statistics, University of Washington, Seattle, Washington USA.*

Yohai, V.J. (1987). High breakdown point and high efficiency robust estimates for regression. *The Annals of Statistics, Vol. 15, pp. 642-656.*

Zou, H. (2006). The adaptive lasso and its oracle properties. *Journal of the American Statistical Association, Vol. 101, pp. 1418-14-29.*

Zou, H., and Li, R. (2008). One-step sparse estimates in nonconcave penalized likelihood models. *The Annals of Statistics, Vol. 36, pp.1509-1533.*

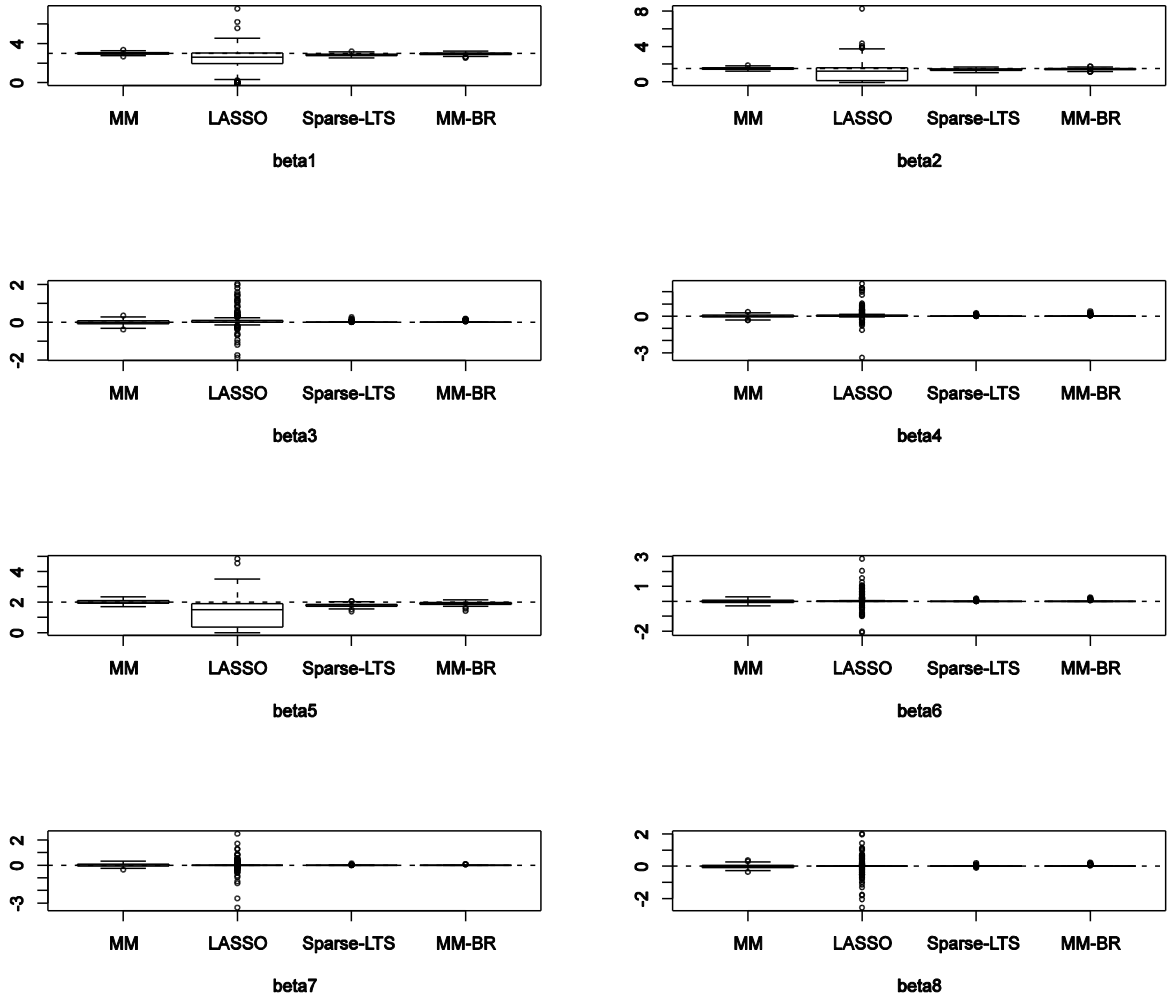

Figure 1: Boxplots of the regression estimates from 200 simulated data sets. The error distribution is Cauchy, n=100, $\varepsilon = 0$ and $\sigma = 0.5$. The horizontal lines correspond to the true parameter values.

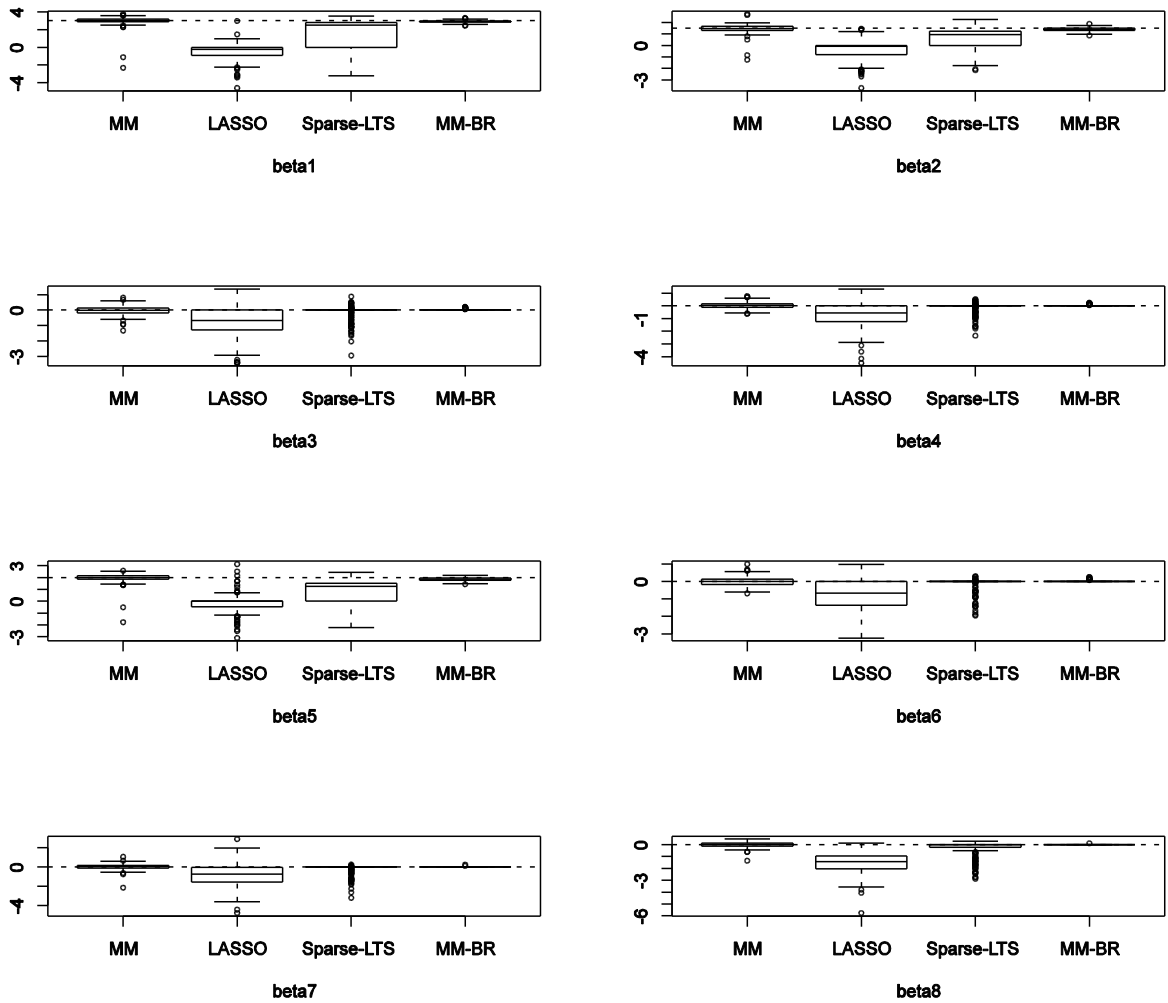

Figure 2: Boxplots of the regression estimates from 200 simulated data sets. The error distribution is Cauchy, n=100, $\varepsilon = 0.3$ and $\sigma = 0.5$. The trimming proportion for the sparse LTS is 0.70. The horizontal lines correspond to the true parameter values.

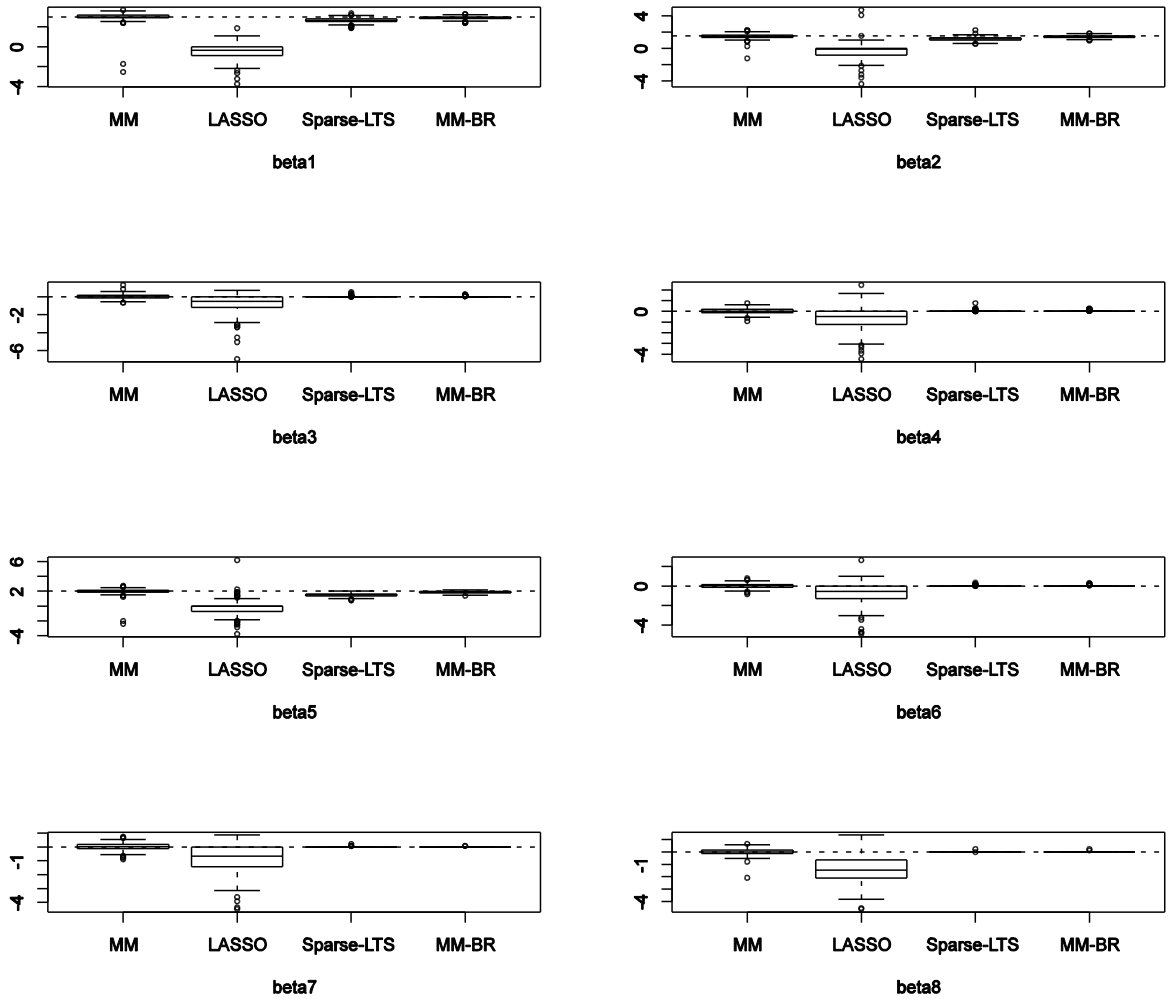

Figure 3: Boxplots of the regression estimates from 200 simulated data sets. The error distribution is Cauchy, n=100, $\varepsilon = 0.3$ and $\sigma = 0.5$. The trimming proportion for the sparse LTS is 0.65. The horizontal lines correspond to the true parameter values.

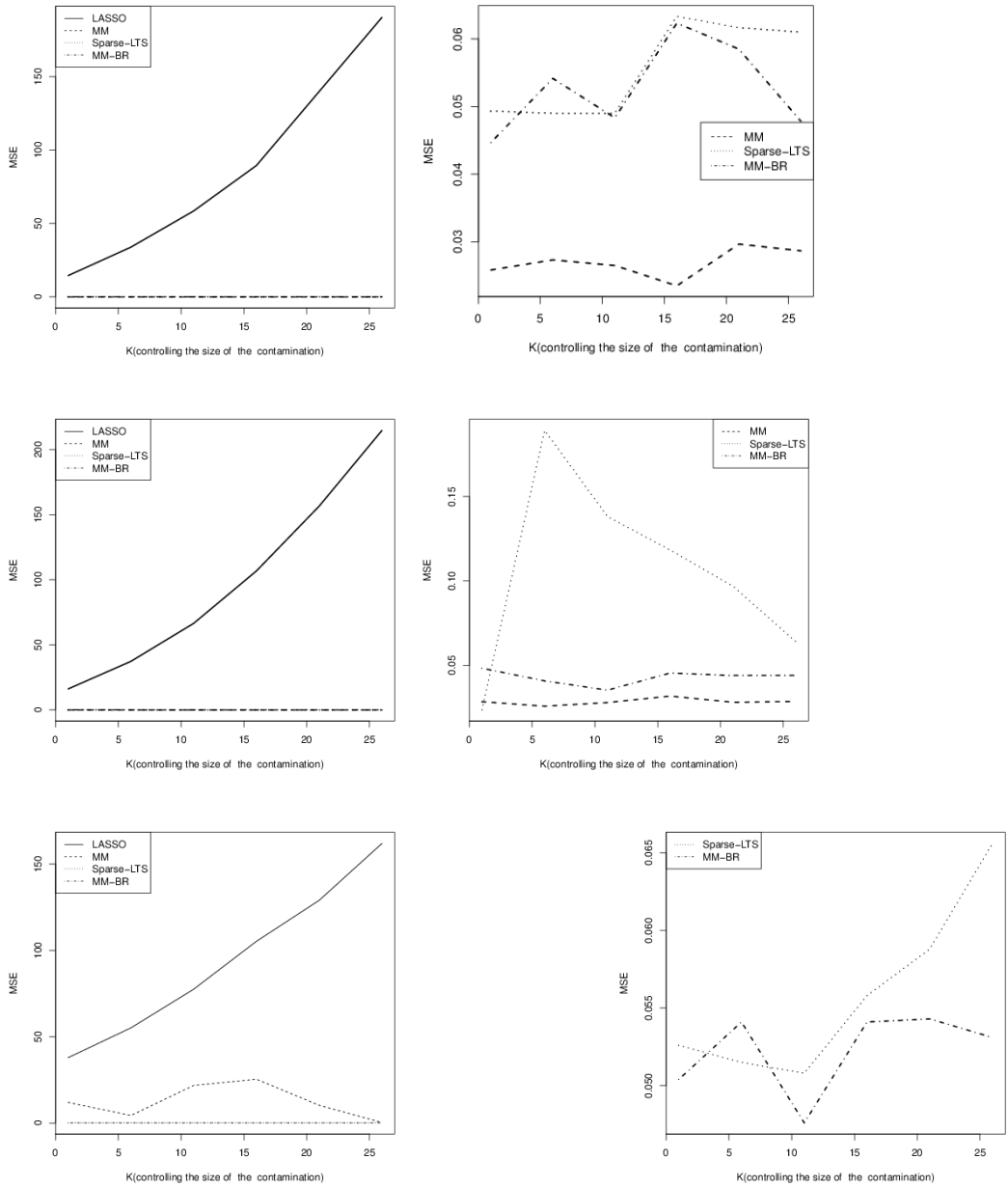

Figure 4: MSE plots for the simulation configuration given in *Case 3*. First row: $p=8, \varepsilon=0.1$. Second row: $p=8, \varepsilon=0.2$. Third row: $p=50, \varepsilon=0.2$.

Table 1: Simulation results for t error ($p = 8$)

$t_1$ error ($\sigma = 0.5$)

| n | $\varepsilon$ | method | No. of zeros | | MeanMSE | MedianMSE |
|---|---|---|---|---|---|---|
| | | | Correct | Incorrect | | |
| | 0.0 | LASSO | 3.08 | 0.59 | 6.11 | 2.27 |
| | | Sparse LTS | 4.05 | 0.00 | 0.19 | 0.16 |
| | | MM-BR | 4.78 | 0.00 | 0.16 | 0.09 |
| | 0.1 | LASSO | 2.70 | 2.36 | 29.94 | 29.20 |
| | | Sparse LTS | 3.58 | 0.00 | 0.16 | 0.12 |
| | | MM-BR | 4.69 | 0.00 | 0.19 | 0.10 |
| | 0.2 | LASSO | 2.07 | 2.10 | 40.64 | 40.39 |
| | | Sparse LTS | 3.82 | 0.01 | 0.55 | 0.38 |
| | | MM-BR | 4.54 | 0.00 | 0.18 | 0.14 |
| | 0.0 | LASSO | 2.98 | 0.69 | 6.00 | 2.37 |
| | | Sparse LTS | 4.48 | 0.00 | 0.12 | 0.11 |
| | | MM-BR | 4.90 | 0.00 | 0.06 | 0.04 |
| | 0.1 | LASSO | 2.46 | 2.24 | 30.21 | 29.60 |
| | | Sparse LTS | 4.01 | 0.00 | 0.07 | 0.06 |
| | | MM-BR | 4.51 | 0.00 | 0.05 | 0.04 |
| | 0.2 | LASSO | 1.77 | 1.97 | 39.95 | 38.97 |
| | | Sparse LTS | 4.37 | 0.00 | 0.25 | 0.18 |
| | | MM-BR | 4.65 | 0.00 | 0.08 | 0.06 |
| | 0.0 | LASSO | 3.06 | 0.49 | 4.62 | 1.77 |
| | | Sparse LTS | 4.77 | 0.00 | 0.11 | 0.10 |
| | | MM-BR | 4.90 | 0.00 | 0.03 | 0.02 |
| | 0.1 | LASSO | 1.81 | 1.92 | 31.32 | 31.49 |
| | | Sparse LTS | 4.47 | 0.00 | 0.06 | 0.05 |
| | | MM-BR | 4.99 | 0.00 | 0.02 | 0.02 |
| | 0.2 | LASSO | 1.22 | 1.84 | 40.88 | 41.01 |
| | | Sparse LTS | 4.64 | 0.00 | 0.18 | 0.16 |
| | | MM-BR | 4.99 | 0.00 | 0.04 | 0.03 |

$t_3$ error ($\sigma = 0.5$)

| | 0.0 | LASSO | 2.48 | 0.00 | 0.10 | 0.07 |
|---|---|---|---|---|---|---|
| | | Sparse LTS | 4.39 | 0.00 | 0.17 | 0.14 |
| | | MM-BR | 4.65 | 0.00 | 0.17 | 0.11 |
| | 0.1 | LASSO | 2.33 | 2.41 | 29.75 | 28.77 |
| | | Sparse LTS | 4.09 | 0.00 | 0.11 | 0.09 |
| | | MM-BR | 4.75 | 0.00 | 0.15 | 0.12 |

| | | | No. of zeros | | | |
|---|---|---|---|---|---|---|
| | | | Correct | Incorrect | MeanMSE | MedianMSE |
| | 0.2 | LASSO | 1.52 | 2.02 | 39.06 | 38.75 |
| | | Sparse LTS | 4.47 | 0.00 | 0.18 | 0.16 |
| | | MM-BR | 4.84 | 0.00 | 0.10 | 0.07 |
| | 0.0 | LASSO | 2.82 | 0.00 | 0.05 | 0.04 |
| | | Sparse LTS | 4.72 | 0.00 | 0.13 | 0.11 |
| | | MM-BR | 4.87 | 0.00 | 0.05 | 0.03 |
| | 0.1 | LASSO | 1.29 | 1.93 | 31.46 | 31.30 |
| | | Sparse LTS | 4.61 | 0.00 | 0.05 | 0.04 |
| | | MM-BR | 4.86 | 0.00 | 0.06 | 0.04 |
| | 0.2 | LASSO | 0.83 | 1.82 | 39.95 | 40.20 |
| | | Sparse LTS | 4.78 | 0.00 | 0.16 | 0.15 |
| | | MM-BR | 4.94 | 0.00 | 0.05 | 0.04 |
| | 0.0 | LASSO | 2.63 | 0.00 | 0.02 | 0.02 |
| | | Sparse LTS | 4.95 | 0.00 | 0.11 | 0.11 |
| | | MM-BR | 4.98 | 0.00 | 0.03 | 0.02 |
| | 0.1 | LASSO | 0.75 | 1.63 | 31.45 | 31.13 |
| | | Sparse LTS | 4.57 | 0.00 | 0.04 | 0.04 |
| | | MM-BR | 4.71 | 0.00 | 0.03 | 0.02 |
| | 0.2 | LASSO | 0.52 | 1.80 | 40.61 | 40.53 |
| | | Sparse LTS | 4.89 | 0.00 | 0.16 | 0.15 |
| | | MM-BR | 4.98 | 0.00 | 0.03 | 0.02 |

Table 1 (Continued)

$t_3$ error ($\sigma = 1$)

| | | | No. of zeros | | | |
|---|---|---|---|---|---|---|
| $n$ | $\varepsilon$ | method | Correct | Incorrect | MeanMSE | MedianMSE |
| | 0.0 | LASSO | 2.88 | 0.00 | 0.38 | 0.28 |
| | | Sparse LTS | 3.62 | 0.00 | 0.41 | 0.30 |
| | | MM-BR | 4.36 | 0.03 | 0.65 | 0.35 |
| | 0.1 | LASSO | 2.23 | 2.31 | 30.07 | 29.36 |
| | | Sparse LTS | 3.09 | 0.00 | 0.30 | 0.26 |
| | | MM-BR | 4.48 | 0.00 | 0.39 | 0.32 |
| | 0.2 | LASSO | 1.75 | 1.81 | 41.05 | 40.56 |
| | | Sparse LTS | 3.84 | 0.00 | 0.39 | 0.32 |
| | | MM-BR | 4.50 | 0.01 | 0.58 | 0.36 |
| | 0.0 | LASSO | 2.79 | 0.00 | 0.18 | 0.14 |
| | | Sparse LTS | 4.19 | 0.00 | 0.22 | 0.21 |
| | | MM-BR | 4.55 | 0.00 | 0.22 | 0.18 |
| | 0.1 | LASSO | 1.24 | 2.10 | 31.06 | 30.59 |

| | | | | | |
|---|---|---|---|---|---|
| | | Sparse LTS | 3.58 | 0.00 | 0.16 | 0.13 |
| | | MM-BR | 4.56 | 0.00 | 0.23 | 0.18 |
| | 0.2 | LASSO | 0.81 | 1.79 | 40.11 | 39.58 |
| | | Sparse LTS | 4.28 | 0.00 | 0.22 | 0.19 |
| | | MM-BR | 4.80 | 0.00 | 0.16 | 0.13 |
| | 0.0 | LASSO | 2.69 | 0.00 | 0.09 | 0.08 |
| | | Sparse LTS | 4.41 | 0.00 | 0.15 | 0.14 |
| | | MM-BR | 4.74 | 0.00 | 0.10 | 0.09 |
| | 0.1 | LASSO | 0.86 | 1.60 | 31.35 | 30.87 |
| | | Sparse LTS | 3.88 | 0.00 | 0.07 | 0.06 |
| | | MM-BR | 4.56 | 0.00 | 0.10 | 0.08 |
| | 0.2 | LASSO | 0.43 | 1.70 | 40.63 | 40.92 |
| | | Sparse LTS | 4.62 | 0.00 | 0.21 | 0.18 |
| | | MM-BR | 4.85 | 0.00 | 0.10 | 0.07 |

Table 2: Simulation results for t error ($p = 50$)

$t_1$ error ($\sigma = 0.5$)

| n | $\varepsilon$ | method | No. of zeros | | MeanMSE | MedianMSE |
|---|---|---|---|---|---|---|
| | | | Correct | Incorrect | | |
| | 0.0 | LASSO | 33.67 | 1.68 | 27.64 | 7.92 |
| | | Sparse LTS | 30.46 | 0.00 | 0.30 | 0.30 |
| | | MM-BR | 39.87 | 0.10 | 1.28 | 1.20 |
| | 0.1 | LASSO | 17.55 | 6.77 | 1049.43 | 1012.33 |
| | | Sparse LTS | 29.87 | 0.00 | 0.38 | 0.30 |
| | | MM-BR | 39.87 | 0.06 | 1.99 | 1.58 |
| | 0.2 | LASSO | 17.52 | 6.48 | 1303.81 | 1084.11 |
| | | Sparse LTS | 33.77 | 0.00 | 0.66 | 0.48 |
| | | MM-BR | 39.73 | 0.26 | 4.88 | 4.34 |
| | 0.0 | LASSO | 32.73 | 1.82 | 19.97 | 8.53 |
| | | Sparse LTS | 35.16 | 0.00 | 0.06 | 0.04 |
| | | MM-BR | 40.00 | 0.00 | 0.05 | 0.04 |
| | 0.1 | LASSO | 9.27 | 4.91 | 1004.57 | 926.14 |
| | | Sparse LTS | 34.32 | 0.00 | 0.09 | 0.09 |
| | | MM-BR | 39.89 | 0.00 | 0.07 | 0.05 |
| | 0.2 | LASSO | 10.11 | 5.56 | 1057.15 | 983.24 |
| | | Sparse LTS | 36.77 | 0.00 | 0.18 | 0.17 |
| | | MM-BR | 39.66 | 0.00 | 0.52 | 0.48 |
| | 0.0 | LASSO | 32.33 | 2.02 | 22.31 | 5.99 |
| | | Sparse LTS | 37.32 | 0.00 | 0.04 | 0.04 |
| | | MM-BR | 40.00 | 0.00 | 0.03 | 0.02 |
| | 0.1 | LASSO | 7.46 | 4.83 | 979.22 | 892.55 |
| | | Sparse LTS | 35.61 | 0.00 | 0.05 | 0.05 |
| | | MM-BR | 40.00 | 0.00 | 0.03 | 0.03 |
| | 0.2 | LASSO | 6.83 | 4.45 | 1037.64 | 964.93 |
| | | Sparse LTS | 36.88 | 0.00 | 0.11 | 0.10 |
| | | MM-BR | 39.49 | 0.00 | 0.24 | 0.22 |
| $t_3$ error ($\sigma = 0.5$) | | | | | | |
| | 0.0 | LASSO | 34.02 | 0.00 | 0.60 | 0.54 |
| | | Sparse LTS | 33.52 | 0.00 | 0.19 | 0.16 |
| | | MM-BR | 39.70 | 1.20 | 12.71 | 0.20 |
| | 0.1 | LASSO | 16.39 | 6.58 | 987.21 | 975.66 |
| | | Sparse LTS | 31.64 | 0.00 | 0.20 | 0.17 |
| | | MM-BR | 39.81 | 0.01 | 0.61 | 0.17 |
| | 0.2 | LASSO | 16.22 | 6.37 | 1096.05 | 1086.52 |
| | | Sparse LTS | 33.82 | 0.00 | 0.18 | 0.16 |
| | | MM-BR | 39.95 | 0.12 | 3.33 | 2.64 |
| | 0.0 | LASSO | 33.34 | 0.00 | 0.05 | 0.05 |
| | | Sparse LTS | 37.87 | 0.00 | 0.05 | 0.05 |
| | | MM-BR | 40.00 | 0.00 | 0.06 | 0.05 |
| | 0.1 | LASSO | 7.63 | 5.35 | 899.78 | 901.36 |
| | | Sparse LTS | 35.95 | 0.00 | 0.04 | 0.04 |
| | | MM-BR | 39.51 | 0.00 | 0.06 | 0.05 |
| | 0.2 | LASSO | 8.36 | 5.10 | 973.52 | 971.38 |
| | | Sparse LTS | 36.73 | 0.00 | 0.04 | 0.04 |
| | | MM-BR | 38.96 | 0.00 | 0.04 | 0.04 |
| | 0.0 | LASSO | 33.72 | 0.00 | 0.03 | 0.03 |
| | | Sparse LTS | 39.47 | 0.00 | 0.04 | 0.04 |
| | | MM-BR | 39.69 | 0.00 | 0.03 | 0.03 |
| | 0.1 | LASSO | 4.58 | 4.81 | 869.82 | 870.67 |
| | | Sparse LTS | 37.40 | 0.00 | 0.02 | 0.02 |
| | | MM-BR | 39.82 | 0.00 | 0.02 | 0.02 |
| | 0.2 | LASSO | 5.35 | 4.64 | 951.69 | 952.17 |
| | | Sparse LTS | 38.14 | 0.00 | 0.03 | 0.03 |
| | | MM-BR | 40.00 | 0.00 | 0.09 | 0.08 |

Table 2 (Continued)

$t_3$ error ($\sigma = 1$)

| n | $\varepsilon$ | method | No. of zeros Correct | Incorrect | MeanMSE | MedianMSE |
|---|---|---|---|---|---|---|
| | 0.0 | LASSO | 33.81 | 0.00 | 0.82 | 0.68 |
| | | Sparse LTS | 27.26 | 0.00 | 1.05 | 0.88 |
| | | MM-BR | 39.16 | 1.44 | 15.58 | 0.90 |
| | 0.1 | LASSO | 15.97 | 6.48 | 998.62 | 1002.58 |
| | | Sparse LTS | 24.79 | 0.00 | 0.91 | 0.75 |
| | | MM-BR | 39.50 | 0.01 | 1.16 | 0.78 |
| | 0.2 | LASSO | 16.64 | 6.24 | 1108.26 | 1101.34 |
| | | Sparse LTS | 29.99 | 0.00 | 0.75 | 0.66 |
| | | MM-BR | 39.93 | 0.18 | 3.80 | 2.89 |
| | 0.0 | LASSO | 32.73 | 0.00 | 0.21 | 0.20 |
| | | Sparse LTS | 35.05 | 0.00 | 0.17 | 0.17 |
| | | MM-BR | 39.80 | 0.00 | 0.25 | 0.20 |
| | 0.1 | LASSO | 7.57 | 5.12 | 904.06 | 902.78 |
| | | Sparse LTS | 34.45 | 0.00 | 0.16 | 0.15 |
| | | MM-BR | 39.78 | 0.00 | 0.19 | 0.18 |
| | 0.2 | LASSO | 8.13 | 4.74 | 975.08 | 976.98 |
| | | Sparse LTS | 34.90 | 0.00 | 0.20 | 0.18 |
| | | MM-BR | 39.96 | 0.00 | 0.33 | 0.32 |
| | 0.0 | LASSO | 32.54 | 0.00 | 0.14 | 0.12 |
| | | Sparse LTS | 34.82 | 0.00 | 0.10 | 0.09 |
| | | MM-BR | 39.93 | 0.00 | 0.13 | 0.12 |
| | 0.1 | LASSO | 4.40 | 4.51 | 874.28 | 874.63 |
| | | Sparse LTS | 36.66 | 0.00 | 0.10 | 0.10 |
| | | MM-BR | 39.98 | 0.00 | 0.12 | 0.11 |
| | 0.2 | LASSO | 5.21 | 4.37 | 947.2 | 944.8 |
| | | Sparse LTS | 36.58 | 0.00 | 0.11 | 0.11 |
| | | MM-BR | 39.99 | 0.00 | 0.12 | 0.12 |

Table 3: Simulation results for normal error ($p = 8$)

**Vertical outliers ($\sigma = 0.5$)**

| n | $\varepsilon$ | method | No. of zeros Correct | Incorrect | MeanMSE | MedianMSE |
|---|---|---|---|---|---|---|
| | 0.0 | LASSO | 3.41 | 0.00 | 0.04 | 0.04 |
| | | Sparse LTS | 4.55 | 0.00 | 0.17 | 0.14 |
| | | MM-BR | 4.48 | 0.03 | 0.38 | 0.12 |
| | 0.1 | LASSO | 2.57 | 0.11 | 2.38 | 2.16 |
| | | Sparse LTS | 4.35 | 0.00 | 0.12 | 0.10 |
| | | MM-BR | 4.65 | 0.00 | 0.14 | 0.10 |
| | 0.2 | LASSO | 2.78 | 0.35 | 4.29 | 3.88 |
| | | Sparse LTS | 4.46 | 0.00 | 0.10 | 0.09 |
| | | MM-BR | 4.74 | 0.00 | 0.09 | 0.07 |
| | 0.0 | LASSO | 3.87 | 0.00 | 0.02 | 0.02 |
| | | Sparse LTS | 4.83 | 0.00 | 0.12 | 0.11 |
| | | MM-BR | 4.83 | 0.00 | 0.06 | 0.04 |
| | 0.1 | LASSO | 2.44 | 0.00 | 1.15 | 1.07 |
| | | Sparse LTS | 4.78 | 0.00 | 0.10 | 0.10 |
| | | MM-BR | 4.87 | 0.00 | 0.05 | 0.04 |
| | 0.2 | LASSO | 2.57 | 0.07 | 2.12 | 1.91 |
| | | Sparse LTS | 4.84 | 0.00 | 0.08 | 0.07 |
| | | MM-BR | 4.94 | 0.00 | 0.04 | 0.03 |
| | 0.0 | LASSO | 3.37 | 0.00 | 0.01 | 0.01 |
| | | Sparse LTS | 4.97 | 0.00 | 0.12 | 0.11 |
| | | MM-BR | 4.96 | 0.00 | 0.03 | 0.02 |
| | 0.1 | LASSO | 2.73 | 0.00 | 0.47 | 0.45 |
| | | Sparse LTS | 4.96 | 0.00 | 0.08 | 0.08 |
| | | MM-BR | 4.94 | 0.00 | 0.03 | 0.02 |
| | 0.2 | LASSO | 2.29 | 0.00 | 0.98 | 0.94 |
| | | Sparse LTS | 4.83 | 0.00 | 0.06 | 0.06 |
| | | MM-BR | 4.98 | 0.00 | 0.02 | 0.02 |

**Vertical outliers ($\sigma = 1$)**

| n | $\varepsilon$ | method | Correct | Incorrect | MeanMSE | MedianMSE |
|---|---|---|---|---|---|---|
| | 0.0 | LASSO | 3.74 | 0.00 | 0.14 | 0.12 |
| | | Sparse LTS | 4.01 | 0.00 | 0.31 | 0.24 |
| | | MM-BR | 4.35 | 0.00 | 0.45 | 0.33 |
| | 0.1 | LASSO | 3.16 | 0.76 | 7.52 | 6.54 |
| | | Sparse LTS | 3.80 | 0.00 | 0.26 | 0.19 |
| | | MM-BR | 4.49 | 0.00 | 0.39 | 0.26 |
| | 0.2 | LASSO | 2.80 | 0.00 | 0.82 | 0.50 |
| | | Sparse LTS | 3.73 | 0.00 | 0.23 | 0.19 |

|   |     | method     | Correct | Incorrect | MeanMSE | MedianMSE |
|---|-----|------------|---------|-----------|---------|-----------|
|   |     | MM-BR      | 4.53    | 0.00      | 0.39    | 0.3       |
|   | 0.0 | LASSO      | 3.57    | 0.00      | 0.17    | 0.14      |
|   |     | Sparse LTS | 4.39    | 0.00      | 0.17    | 0.14      |
|   |     | MM-BR      | 4.46    | 0.00      | 0.21    | 0.16      |
|   | 0.1 | LASSO      | 2.67    | 0.31      | 4.15    | 3.84      |
|   |     | Sparse LTS | 4.30    | 0.00      | 0.13    | 0.12      |
|   |     | MM-BR      | 4.55    | 0.00      | 0.18    | 0.14      |
|   | 0.2 | LASSO      | 2.32    | 0.00      | 0.17    | 0.12      |
|   |     | Sparse LTS | 3.97    | 0.00      | 0.11    | 0.10      |
|   |     | MM-BR      | 4.66    | 0.00      | 0.16    | 0.12      |
|   | 0.0 | LASSO      | 3.19    | 0.00      | 0.05    | 0.04      |
|   |     | Sparse LTS | 4.60    | 0.00      | 0.15    | 0.14      |
|   |     | MM-BR      | 4.72    | 0.00      | 0.12    | 0.10      |
|   | 0.1 | LASSO      | 2.67    | 0.06      | 1.98    | 1.65      |
|   |     | Sparse LTS | 4.52    | 0.00      | 0.10    | 0.10      |
|   |     | MM-BR      | 4.72    | 0.00      | 0.11    | 0.09      |
|   | 0.2 | LASSO      | 2.72    | 0.19      | 3.46    | 2.98      |
|   |     | Sparse LTS | 4.47    | 0.00      | 0.08    | 0.07      |
|   |     | MM-BR      | 4.83    | 0.00      | 0.08    | 0.07      |

Table 3 (Continued)

Leverage points ($\sigma = 0.5$)

| $n$ | $\varepsilon$ | method     | No. of zeros |           | MeanMSE | MedianMSE |
|-----|---------------|------------|--------------|-----------|---------|-----------|
|     |               |            | Correct      | Incorrect |         |           |
|     | 0.1           | LASSO      | 2.37         | 2.32      | 29.46   | 29.01     |
|     |               | Sparse LTS | 3.97         | 0.00      | 0.09    | 0.08      |
|     |               | MM-BR      | 4.67         | 0.00      | 0.14    | 0.10      |
|     | 0.2           | LASSO      | 1.54         | 1.85      | 40.33   | 40.02     |
|     |               | Sparse LTS | 4.62         | 0.00      | 0.19    | 0.16      |
|     |               | MM-BR      | 4.80         | 0.00      | 0.10    | 0.08      |
|     | 0.1           | LASSO      | 1.4 0        | 1.92      | 30.93   | 30.52     |
|     |               | Sparse LTS | 4.53         | 0.00      | 0.05    | 0.05      |
|     |               | MM-BR      | 4.89         | 0.00      | 0.05    | 0.04      |
|     | 0.2           | LASSO      | 1.36         | 1.73      | 32.27   | 32.76     |
|     |               | Sparse LTS | 4.83         | 0.00      | 0.13    | 0.13      |
|     |               | MM-BR      | 4.89         | 0.00      | 0.05    | 0.04      |
|     | 0.1           | LASSO      | 0.61         | 1.49      | 31.34   | 31.31     |
|     |               | Sparse LTS | 4.78         | 0.00      | 0.04    | 0.03      |
|     |               | MM-BR      | 4.99         | 0.00      | 0.03    | 0.02      |
|     | 0.2           | LASSO      | 0.46         | 1.74      | 40.33   | 40.3      |
| | | Sparse LTS | 4.97 | 0.00 | 0.12 | 0.12 |
| | | MM-BR | 4.97 | 0.00 | 0.02 | 0.01 |

Leverage points ($\sigma = 1$)

| | | | | | | |
|---|---|---|---|---|---|---|
| | 0.1 | LASSO | 2.25 | 2.29 | 29.96 | 28.91 |
| | | Sparse LTS | 3.40 | 0.00 | 0.23 | 0.17 |
| | | MM-BR | 4.24 | 0.00 | 0.40 | 0.30 |
| | 0.2 | LASSO | 1.48 | 2.07 | 40.86 | 40.16 |
| | | Sparse LTS | 4.16 | 0.00 | 0.23 | 0.20 |
| | | MM-BR | 4.50 | 0.00 | 0.31 | 0.24 |
| | 0.1 | LASSO | 1.30 | 1.85 | 30.71 | 30.59 |
| | | Sparse LTS | 3.74 | 0.00 | 0.10 | 0.08 |
| | | MM-BR | 4.56 | 0.00 | 0.20 | 0.16 |
| | 0.2 | LASSO | 1.06 | 1.67 | 40.92 | 40.86 |
| | | Sparse LTS | 4.51 | 0.00 | 0.18 | 0.16 |
| | | MM-BR | 4.63 | 0.00 | 0.16 | 0.13 |
| | 0.1 | LASSO | 0.65 | 1.51 | 31.15 | 30.97 |
| | | Sparse LTS | 4.05 | 0.00 | 0.05 | 0.05 |
| | | MM-BR | 4.77 | 0.00 | 0.08 | 0.06 |
| | 0.2 | LASSO | 0.36 | 1.74 | 40.38 | 40.03 |
| | | Sparse LTS | 4.73 | 0.00 | 0.15 | 0.15 |
| | | MM-BR | 4.77 | 0.00 | 0.08 | 0.07 |

Table 4: Simulation results for normal error ($p = 50$)

Vertical outliers ($\sigma = 0.5$)

| | | | No. of zeros | | | |
|---|---|---|---|---|---|---|
| n | $\varepsilon$ | method | Correct | Incorrect | MeanMSE | MedianMSE |
| | 0.0 | LASSO | 34.36 | 0.00 | 0.10 | 0.10 |
| | | Sparse LTS | 35.55 | 0.00 | 0.15 | 0.13 |
| | | MM-BR | 39.83 | 2.60 | 27.19 | 0.22 |
| | 0.1 | LASSO | 30.60 | 0.00 | 4.52 | 4.16 |
| | | Sparse LTS | 35.17 | 0.00 | 0.13 | 0.12 |
| | | MM-BR | 39.89 | 0.20 | 2.24 | 0.14 |
| | 0.2 | LASSO | 31.16 | 0.11 | 7.66 | 7.45 |
| | | Sparse LTS | 34.56 | 0.00 | 0.10 | 0.09 |
| | | MM-BR | 40.00 | 0.00 | 0.15 | 0.14 |
| | 0.0 | LASSO | 35.15 | 0.00 | 0.06 | 0.06 |
| | | Sparse LTS | 39.17 | 0.00 | 0.04 | 0.04 |
| | | MM-BR | 40.00 | 0.00 | 0.07 | 0.06 |
| | 0.1 | LASSO | 30.26 | 0.00 | 1.13 | 1.15 |
| | | Sparse LTS | 38.57 | 0.00 | 0.03 | 0.03 |
| | | MM-BR | 40.00 | 0.00 | 0.06 | 0.06 |
| | 0.2 | LASSO | 29.20 | 0.00 | 2.42 | 2.? |
| | | Sparse LTS | 37.67 | 0.00 | 0.03 | 0.03 |
| | | MM-BR | 40.00 | 0.00 | 0.05 | 0.04 |
| | 0.0 | LASSO | 35.29 | 0.00 | 0.03 | 0.02 |
| | | Sparse LTS | 39.80 | 0.00 | 0.03 | 0.03 |
| | | MM-BR | 40.00 | 0.00 | 0.04 | 0.04 |
| | 0.1 | LASSO | 31.03 | 0.00 | 0.67 | 0.66 |
| | | Sparse LTS | 39.63 | 0.00 | 0.02 | 0.02 |
| | | MM-BR | 40.00 | 0.00 | 0.04 | 0.03 |
| | 0.2 | LASSO | 29.21 | 0.00 | 1.36 | 1.31 |
| | | Sparse LTS | 39.16 | 0.00 | 0.02 | 0.02 |
| | | MM-BR | 40.00 | 0.00 | 0.02 | 0.02 |
| Vertical outliers ($\sigma = 1$) | | | | | | |
| | 0.0 | LASSO | 33.48 | 0.00 | 0.28 | 0.27 |
| | | Sparse LTS | 29.32 | 0.00 | 0.95 | 0.83 |
| | | MM-BR | 39.17 | 2.67 | 28.41 | 0.97 |
| | 0.1 | LASSO | 32.33 | 0.78 | 14.41 | 14.04 |
| | | Sparse LTS | 29.48 | 0.00 | 0.57 | 0.48 |
| | | MM-BR | 39.27 | 0.00 | 0.59 | 0.51 |
| | 0.2 | LASSO | 32.98 | 1.90 | 23.54 | 21.52 |
| | | Sparse LTS | 29.48 | 0.00 | 0.48 | 0.38 |
| | | MM-BR | 40.00 | 0.00 | 0.83 | 0.78 |
| | 0.0 | LASSO | 34.13 | 0.00 | 0.12 | 0.12 |
| | | Sparse LTS | 36.75 | 0.00 | 0.15 | 0.15 |
| | | MM-BR | 39.65 | 0.00 | 0.27 | 0.24 |
| | 0.1 | LASSO | 33.70 | 0.00 | 4.01 | 4.05 |
| | | Sparse LTS | 36.70 | 0.00 | 0.13 | 0.13 |
| | | MM-BR | 39.70 | 0.00 | 0.27 | 0.22 |
| | 0.2 | LASSO | 29.70 | 0.25 | 8.78 | 8.66 |
| | | Sparse LTS | 37.60 | 0.00 | 0.12 | 0.12 |
| | | MM-BR | 39.90 | 0.00 | 0.20 | 0.20 |
| | 0.0 | LASSO | 34.02 | 0.00 | 0.10 | 0.11 |
| | | Sparse LTS | 35.80 | 0.00 | 0.13 | 0.11 |
| | | MM-BR | 39.95 | 0.00 | 0.18 | 0.16 |
| | 0.1 | LASSO | 30.75 | 0.00 | 2.70 | 2.54 |

|  |  | Sparse LTS | 37.05 | 0.00 | 0.08 | 0.07 |
|  |  | MM-BR | 39.95 | 0.00 | 0.14 | 0.16 |
|  | 0.2 | LASSO | 29.05 | 0.10 | 5.48 | 4.82 |
|  |  | Sparse LTS | 37.75 | 0.00 | 0.07 | 0.06 |
|  |  | MM-BR | 40.00 | 0.00 | 0.09 | 0.08 |

Table 4 (Continued)

| Leverage points ($\sigma = 0.5$) | | | | | | |
|---|---|---|---|---|---|---|
| | | | No. of zeros | | | |
| $n$ | $\varepsilon$ | method | Correct | Incorrect | MeanMSE | MedianMSE |
|  | 0.1 | LASSO | 16.34 | 6.88 | 979.28 | 976.09 |
|  |  | Sparse LTS | 34.20 | 0.00 | 0.13 | 0.12 |
|  |  | MM-BR | 39.86 | 0.20 | 2.42 | 0.16 |
|  | 0.2 | LASSO | 16.33 | 6.19 | 1090.77 | 1085.17 |
|  |  | Sparse LTS | 35.27 | 0.00 | 0.09 | 0.08 |
|  |  | MM-BR | 40.00 | 0.01 | 2.57 | 2.18 |
|  | 0.1 | LASSO | 7.88 | 5.24 | 892.3 | 895.13 |
|  |  | Sparse LTS | 37.13 | 0.00 | 0.03 | 0.03 |
|  |  | MM-BR | 40.00 | 0.00 | 0.06 | 0.05 |
|  | 0.2 | LASSO | 8.08 | 5.11 | 984.03 | 983.28 |
|  |  | Sparse LTS | 38.30 | 0.00 | 0.03 | 0.03 |
|  |  | MM-BR | 40.00 | 0.00 | 0.20 | 0.20 |
|  | 0.1 | LASSO | 4.80 | 4.80 | 871.78 | 873.05 |
|  |  | Sparse LTS | 38.68 | 0.00 | 0.02 | 0.02 |
|  |  | MM-BR | 40.00 | 0.00 | 0.03 | 0.03 |
|  | 0.2 | LASSO | 5.10 | 4.37 | 954.01 | 952.88 |
|  |  | Sparse LTS | 39.47 | 0.00 | 0.02 | 0.02 |
|  |  | MM-BR | 40.00 | 0.00 | 0.07 | 0.07 |
| Leverage points ($\sigma = 1$) | | | | | | |
|  | 0.1 | LASSO | 15.95 | 6.15 | 1009.21 | 1002.58 |
|  |  | Sparse LTS | 25.93 | 0.00 | 0.59 | 0.55 |
|  |  | MM-BR | 39.74 | 0.00 | 0.95 | 0.72 |
|  | 0.2 | LASSO | 16.20 | 6.52 | 1080.11 | 1074.94 |
|  |  | Sparse LTS | 29.58 | 0.00 | 0.46 | 0.40 |
|  |  | MM-BR | 38.56 | 0.03 | 2.77 | 2.37 |
|  | 0.1 | LASSO | 7.35 | 4.95 | 906.58 | 897.05 |
|  |  | Sparse LTS | 36.88 | 0.00 | 0.11 | 0.11 |
|  |  | MM-BR | 39.87 | 0.00 | 0.23 | 0.22 |

|   | 0.2 | LASSO | 8.15 | 5.27 | 981.55 | 980.62 |
|   |   | Sparse LTS | 37.38 | 0.00 | 0.12 | 0.12 |
|   |   | MM-BR | 40.00 | 0.00 | 0.24 | 0.24 |
|   | 0.1 | LASSO | 5.12 | 4.90 | 865.23 | 864.13 |
|   |   | Sparse LTS | 37.78 | 0.00 | 0.07 | 0.07 |
|   |   | MM-BR | 39.93 | 0.00 | 0.13 | 0.12 |
|   | 0.2 | LASSO | 5.38 | 4.27 | 954.93 | 950.60 |
|   |   |   |   |   |   |   |
|   |   | Sparse LTS | 37.42 | 0.00 | 0.07 | 0.06 |
|   |   | MM-BR | 39.82 | 0.00 | 0.09 | 0.08 |

Table 5: Simulation results for $\widetilde{\beta}_{MM-BR}^{(1)}$

| $t_1$ error and $p=8$ | | | | | | |
|---|---|---|---|---|---|---|
|   |   |   | No. of zeros | | | |
| $n$ | $\varepsilon$ | method | Correct | Incorrect | MeanMSE | MedianMSE |
|   | 0.0 | MM-BR (One-Step) | 3.25 | 0.00 | 0.20 | 0.12 |
|   | 0.1 | MM-BR (One-Step) | 3.22 | 0.00 | 0.18 | 0.14 |
|   | 0.2 | MM-BR (One-Step) | 3.09 | 0.00 | 0.28 | 0.20 |
|   | 0.0 | MM-BR (One-Step) | 3.63 | 0.00 | 0.06 | 0.04 |
|   | 0.1 | MM-BR (One-Step) | 3.35 | 0.00 | 0.06 | 0.05 |
|   | 0.2 | MM-BR (One-Step) | 3.42 | 0.00 | 0.09 | 0.07 |
|   | 0.0 | MM-BR (One-Step) | 3.97 | 0.00 | 0.02 | 0.02 |
|   | 0.1 | MM-BR (One-Step) | 3.75 | 0.00 | 0.03 | 0.02 |
|   | 0.2 | MM-BR (One-Step) | 3.59 | 0.00 | 0.04 | 0.03 |

| $t_3$ error and $p=8$ | | | | | | |
|---|---|---|---|---|---|---|
|   |   |   | No. of zeros | | | |
| $n$ | $\varepsilon$ | method | Correct | Incorrect | MeanMSE | MedianMSE |
|   | 0.0 | MM-BR | 3.48 | 0.00 | 0.08 | 0.06 |

|   | ε | method | Correct | Incorrect | MeanMSE | MedianMSE |
|---|---|---|---|---|---|---|
|   |   | (One-Step) |   |   |   |   |
|   | 0.1 | MM-BR (One-Step) | 3.22 | 0.00 | 0.11 | 0.09 |
|   | 0.2 | MM-BR (One-Step) | 3.26 | 0.00 | 0.10 | 0.08 |
|   | 0.0 | MM-BR (One-Step) | 3.79 | 0.00 | 0.03 | 0.03 |
|   | 0.1 | MM-BR (One-Step) | 3.60 | 0.00 | 0.04 | 0.03 |
|   | 0.2 | MM-BR (One-Step) | 3.60 | 0.00 | 0.04 | 0.03 |
|   | 0.0 | MM-BR (One-Step) | 4.21 | 0.00 | 0.01 | 0.01 |
|   | 0.1 | MM-BR (One-Step) | 4.13 | 0.00 | 0.01 | 0.01 |
|   | 0.2 | MM-BR (One-Step) | 3.95 | 0.00 | 0.02 | 0.02 |

$t_1$ error and $p = 50$

|   |   |   | No. of zeros | | | |
|---|---|---|---|---|---|---|
| $n$ | ε | method | Correct | Incorrect | MeanMSE | MedianMSE |
|   | 0.0 | MM-BR (One-Step) | 36.26 | 0.00 | 1.59 | 0.83 |
|   | 0.1 | MM-BR (One-Step) | 32.49 | 0.52 | 19.57 | 20.83 |
|   | 0.2 | MM-BR (One-Step) | 30.02 | 1.36 | 37.25 | 33.39 |
|   | 0.0 | MM-BR (One-Step) | 39.75 | 0.00 | 0.46 | 0.07 |
|   | 0.1 | MM-BR (One-Step) | 39.59 | 0.00 | 0.25 | 0.08 |
|   | 0.2 | MM-BR (One-Step) | 38.22 | 0.04 | 3.02 | 1.82 |
|   | 0.0 | MM-BR (One-Step) | 39.99 | 0.00 | 0.23 | 0.03 |
|   | 0.1 | MM-BR (One-Step) | 39.95 | 0.00 | 0.22 | 0.04 |
|   | 0.2 | MM-BR (One-Step) | 38.79 | 0.01 | 2.41 | 2.44 |

$t_3$ error and $p = 50$

|   |   |   | No. of zeros | | | |
|---|---|---|---|---|---|---|
| $n$ | ε | method | Correct | Incorrect | MeanMSE | MedianMSE |
|   | 0.0 | MM-BR (one-Step) | 36.37 | 0.01 | 1.94 | 0.89 |

| | 0.1 | MM-BR (One-Step) | 32.90 | 0.58 | 17.60 | 15.00 |
| --- | --- | --- | --- | --- | --- | --- |
| | 0.2 | MM-BR (One-Step) | 30.03 | 1.43 | 40.76 | 36.44 |
| | 0.0 | MM-BR (One-Step) | 39.67 | 0.00 | 0.78 | 0.09 |
| | 0.1 | MM-BR (One-Step) | 39.62 | 0.00 | 0.31 | 0.08 |
| | 0.2 | MM-BR (One-Step) | 38.50 | 0.05 | 2.92 | 1.44 |
| | 0.0 | MM-BR (One-Step) | 39.98 | 0.00 | 0.26 | 0.05 |
| | 0.1 | MM-BR (One-Step) | 39.95 | 0.00 | 0.26 | 0.04 |
| | 0.2 | MM-BR (One-Step) | 38.84 | 0.01 | 2.47 | 2.13 |

Table 6: Selected variables for pollution data set

| Method | Selected variables |
|---|---|
| LASSO | (1,2,3,6,7,8,9,14) |
| Sparse LTS | (1,2,3,4,5,6,8,11,14,15) |
| MM-BR ($\gamma = 1$) | (1,2,3,7,8,9,11,13,14,15) |
| MM-BR ($\gamma = 0.7$) | (1,2,3,7,8,9,11,14,15) |
| Bridge ($\gamma = 0.7$) | (1,2,3,6,8,9,14) |

Table 7: Prediction errors for pollution data

| Method | LASSO | Sparse LTS | MM-BR ($\gamma = 1$) | MM-BR ($\gamma = 0.7$) | Bridge ($\gamma = 0.7$) |
|---|---|---|---|---|---|
| Prediction error | 1582.94 | 1487.48 | 1486.67 | 1468.24 | 1553.08 |